\documentclass[twocolumn,aps,showpacs,preprintnumbers,nofootinbib,prd,
superscriptaddress,10pt]{revtex4-1}

\usepackage{lmodern}
\usepackage{amsmath,amssymb}
\usepackage[normalem]{ulem}
\usepackage{textcomp}
\usepackage{hyperref}
\usepackage{bm}
\usepackage{graphicx}
\usepackage{psfrag}
\usepackage{aas_macros}
\usepackage[usenames,dvipsnames]{xcolor}
\usepackage[utf8]{inputenc}
\usepackage{lipsum} 
\usepackage{multirow}
\usepackage{subfigure}
\usepackage{acronym}
\usepackage{orcidlink} 
\usepackage{enumitem}
\usepackage{multirow}
\usepackage{booktabs}

\definecolor{darkgreen}{rgb}{0,0.5,0}

\hypersetup{
    unicode=false,          
    pdftoolbar=true,        
    pdfmenubar=true,        
    pdffitwindow=false,     
    pdfstartview={FitH},    
	pdftitle={Parametrizing the Unknown: },
    pdfauthor={Sumit K et al.},
    pdfsubject={Gravitational Waves},   
    pdfkeywords={gravitational waves, eparameter estimation},
    pdfnewwindow=true,      
    colorlinks=true,       
    linkcolor=red,          
    citecolor=Blue,        
    filecolor=magenta,      
    urlcolor=Blue,           
    linktocpage=true
}

\allowdisplaybreaks[4]

\DeclareFontFamily{OT1}{pzc}{}
\DeclareFontShape{OT1}{pzc}{m}{it}{<-> s * [1.10] pzcmi7t}{}
\DeclareMathAlphabet{\mathpzc}{OT1}{pzc}{m}{it}

\usepackage{xspace}

\newcommand{\imrphenomxphm}{\texttt{IMRPhenomXPHM}\xspace}
\newcommand{\seobnr}{\texttt{SEOBNRv5PHM}\xspace}
\newcommand{\nrsur}{\texttt{NRSur7dq4}\xspace}
\newcommand{\imrphenomxo}{\texttt{IMRPhenomXO4a}\xspace}
\usepackage{standalone}  

\begin{document}
    \newacro{nr}[NR]{Numerical relativity}
    
    \newacro{gw}[GW]{gravitational wave}
    
    \newacro{pe}[PE]{parameter estimation}

    \newacro{psd}[PSD]{Power Spectral Density}

    \newacro{pn}[PN]{Post-Newtonian}

    \newacro{snr}[SNR]{signal-to-noise ratio}

    \newacro{bbh}[BBH]{Binary Black Hole}

    \newacro{bns}[BNS]{Binary Neutron Star}

    \newacro{nsbh}[NSBH]{Neutron Star-Black Hole}

    \newacro{gwosc}[GWOSC]{Gravitational Wave Open Science Center}


\title{Mitigating Systematic Errors in Parameter Estimation of Binary Black Hole Mergers in O1-O3 LIGO-Virgo Data}

\date{\today}

\author{Sumit Kumar \orcidlink{0000-0002-6404-0517}}
\email{s.kumar2@uu.nl}
\affiliation{Institute for Gravitational and Subatomic Physics (GRASP), 
Utrecht University, Princetonplein 1, 3584 CC Utrecht, The Netherlands}
\affiliation{Nikhef -- National Institute for Subatomic Physics, 
Science Park 105, 1098 XG Amsterdam, The Netherlands}
\affiliation{Max-Planck-Institut f{\"u}r Gravitationsphysik (Albert-Einstein-Institut), D-30167 Hannover, Germany}
\affiliation{Leibniz Universit{\"a}t Hannover, D-30167 Hannover, Germany}

\author{Max Melching \orcidlink{0009-0001-4899-9955}}
\affiliation{Max-Planck-Institut f{\"u}r Gravitationsphysik (Albert-Einstein-Institut), D-30167 Hannover, Germany}
\affiliation{Leibniz Universit{\"a}t Hannover, D-30167 Hannover, Germany}
\affiliation{LIGO Laboratory, California Institute of Technology, Pasadena, CA 91125, USA}

\author{Frank Ohme \orcidlink{0000-0003-0493-5607}}
\affiliation{Max-Planck-Institut f{\"u}r Gravitationsphysik (Albert-Einstein-Institut), D-30167 Hannover, Germany}
\affiliation{Leibniz Universit{\"a}t Hannover, D-30167 Hannover, Germany}

\author{Harsh Narola \orcidlink{0000-0001-9161-7919}}
\affiliation{Institute for Gravitational and Subatomic Physics (GRASP), 
Utrecht University, Princetonplein 1, 3584 CC Utrecht, The Netherlands}
\affiliation{Nikhef -- National Institute for Subatomic Physics, 
Science Park 105, 1098 XG Amsterdam, The Netherlands}

\author{Tom Dooney \orcidlink{0009-0008-3914-9554}}
\affiliation{Institute for Gravitational and Subatomic Physics (GRASP), 
Utrecht University, Princetonplein 1, 3584 CC Utrecht, The Netherlands}
\affiliation{Nikhef -- National Institute for Subatomic Physics, 
Science Park 105, 1098 XG Amsterdam, The Netherlands}

\author{Chris Van Den Broeck \orcidlink{0000-0001-6800-4006}}
\affiliation{Institute for Gravitational and Subatomic Physics (GRASP), 
Utrecht University, Princetonplein 1, 3584 CC Utrecht, The Netherlands}
\affiliation{Nikhef -- National Institute for Subatomic Physics, 
Science Park 105, 1098 XG Amsterdam, The Netherlands}

\begin{abstract}
Systematic errors in the \ac{pe} of \ac{gw} mergers can arise from various sources, including waveform systematics, noise mischaracterization, data analysis artifacts, and other unknown factors. In this study, we analyze selected events from the first three observing runs of the LIGO-Virgo-KAGRA (LVK) collaboration. We choose events that have been flagged in various studies as potentially affected by systematic errors. They show either differences in posterior samples across different waveform models or a multimodal distribution in one of the key parameters. 
Here we analyze these events again by using a couple of parametric models, developed in previous works, that incorporate uncertainties in both the phase and amplitude of the \ac{gw} waveform. In this data-driven approach, we apply sufficiently broad priors on the uncertainty parameters to account for potential systematic errors. 
Our findings show that the proposed method effectively reduces systematic errors, even those arising from data artifacts, such as glitches occurring near a signal and the deglitching process in \ac{gw} frame files. We find that previously significant differences in the analyses of either raw or deglitched frame files are mitigated in our uncertainty-aware analyses. Similarly, inconsistent results from different waveform models become much more consistent in our framework. One noteworthy event we examine is GW191109\_010717, which is particularly interesting due to its anti-aligned spin properties. We report that, within our framework, the event still exhibits anti-aligned spin characteristics, but the inference results become consistent across raw and deglitched frame files, as well as across the waveform models used for this event (\imrphenomxphm, \imrphenomxo, and \nrsur). A similar trend is observed for the event GW200129\_065458, which previously yielded a high, but inconsistent precession parameter among different waveform models. In contrast, we observe a non-zero and consistent value of $\chi_{p}=0.60^{+0.31}_{-0.33}, 0.58^{+0.30}_{-0.29}$ and $0.56^{+0.31}_{-0.28}$ for the \imrphenomxphm, \imrphenomxo, and \nrsur waveform models, respectively.
\end{abstract}

\date{\today}

\maketitle
\section{Introduction}
\label{sec:intro}

Since the beginning of \ac{gw} astronomy, the field's focus has gradually expanded from identifying and detecting signals in data to also conducting precision science with them. The \ac{gw} catalogs \citep{gwtc4, gwtc3, Nitz:2021zwj, Olsen:2022pin, Mehta:2023zlk}, as well as loud individual events \citep{LIGOScientific:2025rsn, LIGOScientific:2025wao, LIGOScientific:2020iuh}, are used to infer the properties of the underlying population of compact binary mergers \citep{LIGOScientific:2025pvj, KAGRA:2021duu}, test general relativity \citep{LIGOScientific:2025wao, LIGOScientific:2021sio, LIGOScientific:2026qni, LIGOScientific:2026fcf, LIGOScientific:2026wpt}, constrain the material properties of neutron stars \citep{LIGOScientific:2018cki, Capano:2019eae}, estimate the Hubble constant \citep{LIGOScientific:2021aug, LIGOScientific:2025jau}, among many other applications. In addition, ongoing efforts are exploring potential gravitational lensing signatures of \ac{gw} mergers \citep{LIGOScientific:2025cwb, LIGOScientific:2023bwz}. All these investigations rely on the accuracy of the scientific analysis and the quality of data products derived from the searches and \ac{pe} pipelines. However, we have already started documenting the systematic errors in the analysis of many \ac{gw} merger events \citep{gwtc3, gwtc4, LIGOScientific:2025rsn}. These systematic effects include inconsistencies in inferred source parameters for different waveform approximants \citep{LIGOScientific:2025rsn, gwtc3, gwtc4}, or biases introduced by glitches near the signal \citep{Udall:2024ovp}. The primary sources of these systematic biases can be inaccuracies in waveform models in some regions of parameter space \citep{Dhani:2024jja, Nagar:2023zxh, Samajdar:2018dcx, Samajdar:2019ulq, Owen:2023mid, Gamba:2020wgg, Read:2023hkv}, data analysis artifacts (glitches near the signal \citep{Hourihane:2025vxc, Udall:2025bts}, glitch mitigation schemes \citep{Udall:2024ovp, Hourihane:2022doe, Payne:2022spz, Ghonge:2023ksb}, non-Gaussian and/or non-stationary noise \citep{Kumar:2022tto, Plunkett:2022zmx, Bini:2026kwz, Zackay:2019kkv}, etc.), or missing physical effects \citep{Kumar:2025nwb, Chatziioannou:2014coa, Purrer:2015nkh, CalderonBustillo:2015lrt, London:2017bcn, Cotesta:2018fcv, Kalaghatgi:2019log, Pratten:2020igi, Garcia-Quiros:2020qpx, Krishnendu:2021cyi} in the description of the waveform models.  In the coming years, planned upgrades to the LIGO, Virgo, and KAGRA detectors are expected to enhance the network's sensitivity \citep{KAGRA:2013rdx, aLIGO:2020wna, LIGOScientific:2014pky, Abbott:2016xvh}. As detectors become more sensitive, we expect systematic errors to become increasingly significant. Therefore, it will be essential to address these errors more frequently, especially as they may become comparable to statistical errors for larger numbers of events \citep{Kumar:2025nwb}.

The accuracy requirements of a waveform model in a given parameter space depend on the detector's sensitivity and the signal's loudness \citep{Lindblom:2008cm}. As detector sensitivity improves, waveform models need to achieve correspondingly higher levels of accuracy \citep{Purrer:2019jcp}. A given waveform approximant may suffer from inaccuracies inherent to the waveform-model construction \citep{Kunert:2024xqb, Hu:2022rjq}, uncertainties in the numerical relativity calibration schemes \citep{Pompili:2024yec, Mezzasoma:2026wme}, or the extrapolation of calibration schemes into regions of parameter space where only a few numerical relativity simulations are available \citep{Dhani:2024jja}. In the literature, many schemes are proposed and employed to account for waveform systematic uncertainties \citep{LIGOScientific:2016ebw}. Some of these schemes fall into the category of accounting for systematic differences across waveform models: either by mixing posterior samples from multiple \ac{pe} runs \citep{KAGRA:2021vkt, Ashton:2019leq, Jan:2020bdz} or by assigning appropriate weights to different waveform models during the likelihood call \citep{Ashton:2021cub, Hoy:2022tst, Puecher:2023rxw, PhysRevD.98.124030, PhysRevD.100.024046,  Hoy:2024vpc}. Another category of mitigating scheme is one in which the uncertainties associated with a given waveform model are estimated and incorporated into the \ac{pe} analysis \citep{Pompili:2024yec, Bachhar:2024olc, Kumar:2025nwb, Mezzasoma:2025moh}.

There are additional ways in which a waveform model can be inaccurate, primarily due to the omission of physical or environmental factors in its description. Examples of such scenarios include analyzing a binary merger with eccentricity using a model that does not include it \citep{PhysRevD.109.043037, Divyajyoti:2023rht, Ramos-Buades:2019uvh, Brown:2009ng, Narayan:2023vhm, Guo:2022ehk, Bonino:2022hkj, GilChoi:2022waq, Lee:2025zjx}. Additionally, if the model fails to include specific physical effects, such as higher modes or precession, it can introduce systematic inaccuracies \citep{Chatziioannou:2014coa, Purrer:2015nkh, CalderonBustillo:2015lrt, London:2017bcn, Cotesta:2018fcv, Kalaghatgi:2019log, Pratten:2020igi, Garcia-Quiros:2020qpx, Krishnendu:2021cyi, Divyajyoti:2025cwq}. The impact of these omissions depends on the \ac{snr} of the signal and the location of the source in parameter space. In a recent study, it is argued that not all events need to use the state-of-the-art waveform models, which use higher-order multiple modes and spin-induced quadruple moments \citep{Hoy:2026izf}.

Apart from waveform modelling uncertainties, data analysis artifacts, such as the presence of glitches near the signal, can also affect the \ac{pe} results. Typically, the LIGO-Virgo interferometers encounter glitches at the rate of $\mathcal{O}(1)$ per minute~\cite{KAGRA:2021vkt}. Some of them tend to occur in the vicinity of a gravitational wave signal. Such glitches need to be carefully reconstructed and subtracted from the data before the signal can be analysed; otherwise, their presence may introduce biases in the measurement of the source properties of the signals and in subsequent science cases. For the gravitational wave transient catalogues (GWTC)~\cite{gwtc4, gwtc3}, the glitch mitigation is performed using the BayesWave codebase \citep{Hourihane:2022doe}. The codebase jointly reconstructs the signal and the glitch. It models the signal coherently across the detector network using phenomenological waveforms, while glitches are modelled independently for each detector using a sum of Morlet-Gabor wavelets~\cite{Cornish:2020dwh}.

In this work, we identify the events from the first three observing runs (O1-O3), which are flagged as having potential systematic errors. These events could have waveform systematics or be affected by other data-analysis artifacts that we will discuss in detail. We use a method introduced in \citep{Kumar:2025nwb} to mitigate the systematic errors. In this mitigation scheme, we account for waveform-systematics by introducing amplitude and phase uncertainties in the waveform models. A key feature of the framework used in this work is its general applicability to different types of systematic bias in \ac{gw} \ac{pe}. Though this method is designed to address issues of waveform systematics and missing physical effects, we also find that the framework is useful for addressing certain data analysis artifacts, such as the presence of glitches or potential shortfalls in glitch-mitigation schemes. The posterior distributions of waveform uncertainty parameters also serve as a diagnostics tool for potential systematics, with significant support away from zero indicating unresolved modeling inaccuracies or data-analysis artifacts. This makes the approach broadly applicable for identifying and mitigating systematic uncertainties in current and future \ac{gw} observations.

For this study, we focus on a few selected events from the first three observation runs. This approach allows us to apply our systematic framework to a variety of scenarios and conduct a thorough analysis. Additionally, we use these events to examine and improve our systematic mitigation framework.

The structure of this paper is organized as follows: In section \ref{sec:events_selection}, we briefly discuss the motivation behind the selection of events for this study. In Section \ref{sec:methodology}, we introduce the methodology and provide relevant background information. In Section \ref{sec:wf_differences}, we examine the intrinsic differences among various waveform models within the parameter space relevant to the selected events. We outline the specifics of our analysis and results in Section \ref{sec:events_analysis}. Finally, we summarize our findings and discuss limitations of our method, along with future directions in \ref{sec:summary}.

\section{Selection of the events}
\label{sec:events_selection}
In this section, we list the events selected for this study and the motivation for their inclusion. We will revisit each event in detail in the later sections. The following events are identified as potentially exhibiting waveform systematics in the GWTC-3 \citep{gwtc3} analysis: GW191109\_010717, GW191219\_163120, GW200129\_065458, and GW200208\_222617. Most of these events exhibit inconsistent results across different waveform models in the recovery of a few key parameters, such as chirp mass $(\mathcal{M})$, mass ratio ($q=m_2/m_1;~m_2 \leq m_1$), effective spin parameter ($\chi_\mathrm{eff}$), spin precessing parameter ($\chi_\mathrm{p}$), etc. The events GW191109\_010717 and GW200129\_065458 were affected by a glitch in one of the detectors near the merger time. To mitigate the impact of these noise transients, glitch-mitigation procedures were applied, and new \ac{gw} frame files were generated for subsequent analyses. 
\begin{figure*}
    \centering
    \includegraphics[scale=0.9]{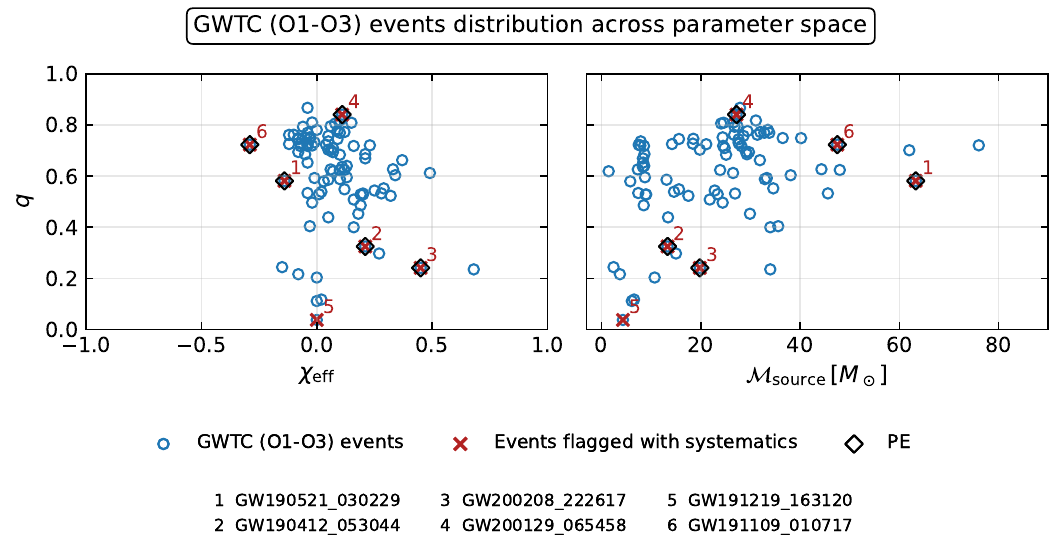}
    \caption{This figure shows the distribution of compact binary mergers from the GWTC catalogs \citep{gwtc21, gwtc3}, which cover the first three observation runs: O1-O3. The blue circles represent all the events across parameter space of effective spin parameter $(\chi_\mathrm{eff})$, mass ratio $(q = \frac{m_2}{m_1}; m_2 \leq m_1)$, and source frame chirp mass $\mathcal{M}_\mathrm{eff}$ in the units of solar mass ($M_{\odot}$). These values are derived from the events table on the \ac{gwosc} portal \citep{LIGOScientific:2019lzm, KAGRA:2023pio} and are estimated based on the default \ac{pe} run listed on each event's page on the portal. The red cross label indicates events flagged with systematic errors across various studies. The numbers near these events are for identification and mapping between two different panels. The black diamond-shaped points represent the events used in the present work to mitigate systematic errors. We exclude GW191219\_163120 because, for the current study, we focus only on binary black hole mergers. }
    \label{fig:scatter_plot_gwtc}
\end{figure*}

The event GW190412\_053044 shows inconsistent estimates for the effective spin parameter between the two waveform models \imrphenomxphm and \seobnr used in the GWTC analysis. GW190521\_030229 is known for bimodal posterior distribution in the recovered mass ratio and chirp mass. Figure~\ref{fig:scatter_plot_gwtc} shows all events from the first three observing runs in the $\chi_\mathrm{eff}-q$ and $\mathcal{M}_\mathrm{source}-q$ planes. We also highlight events flagged for potential systematic errors. 

Among all the events exhibiting systematic errors, GW191219\_163120 has the most asymmetric mass distribution, and is potentially a \ac{nsbh} system. In this work, we focus solely on \ac{bbh} mergers and leave out GW191219\_163120.  These events also serve as useful test cases for validating and improving our systematic-mitigation framework. We will treat \ac{bns} and \ac{nsbh} systems in a follow-up study.

\section{Methodology: Incorporating Waveform Uncertainties}
\label{sec:methodology}
The output of the \ac{gw} detector is a time series, $d(t_i)$, which contains the noise $n(t_i)$ and may contain a transient signal $h(t_i)$ and/or glitch(es) $g(t_i)$. The noise is assumed to be Gaussian and stationary, and is characterized by the \ac{psd}, $S_{n}(f)$, of the detector, which is defined as the Fourier transform of the noise covariance $C_{ij}=\hat{E}[(n(t_i)-\mu)(n(t_j)-\mu)]$, where $\hat{E}$ represents the expectation value and $\mu = \hat{E}[n(t)]$ \citep{LIGOScientific:2019hgc}. The \ac{gw} signal $h(t, \mathbf{\Theta})$ is modeled using waveform approximants calibrated to numerical relativity simulations. Here $t$ denotes time and $\mathbf{\Theta}$ denotes the model parameters. For a typical \ac{gw} signal from a \ac{bbh} merger, these parameters are the component masses of the binary system ($m_{1,2}$), component spins ($\vec{S}_{1,2}$), luminosity distance ($D_L$), inclination angle between the orbital angular momentum and the line of sight ($\iota$), polarization angle ($\psi$), coalescence phase ($\phi$), right ascension (RA), declination (dec), and time of coalescence ($t_c$). For a \ac{bns} system, there are additional parameters such as the tidal deformability parameter $\Lambda_{1,2}$.

In the Bayesian analysis, the posterior probability distribution $p(\mathbf{\Theta} \mid d,I)$ represents the probability distribution of parameters $\mathbf{\Theta}$ given the realization of the data $d$, and is given by the Bayes' theorem,
\begin{equation}
    p(\mathbf{\Theta} \mid d,I) = \frac{\mathcal{L}(d \mid \mathbf{\Theta},I)\pi(\mathbf{\Theta} \mid I)}{p(d \mid I)}, \label{eqn:bayes_theorem}
\end{equation}
where $\mathcal{L}(d \mid \mathbf{\Theta},I)$ denotes the likelihood function for the data realization $d$ given the model with parameters $\mathbf{\Theta}$, $\pi(\mathbf{\Theta}|I)$ is the prior probability distribution of the parameters $\mathbf{\Theta}$, and $p(d \mid I)$ is called marginalized likelihood or Bayesian evidence for the model. The $I$ represents any additional prior knowledge about the system.

The statistical uncertainty $\sigma_{\Theta_i}$ in the parameter $\Theta_i$ can be estimated from the width of the marginalized posterior distribution,
\begin{equation}
    p(\Theta_i \mid d, I) = \int p(\mathbf{\Theta} \mid d, I)~\prod_{j \neq i} d\Theta_{j }.
\end{equation}
 The ensemble-averaged systematic bias $\Delta\Theta_i = \bar{\Theta}_i - \Theta_{i, \mathrm{true}}$ is defined as the difference between the mean recovered value of the parameter over multiple noise realizations, $\bar{\Theta}_{i}$, and the true value of the parameter $\Theta_{i,\mathrm{true}}$. The systematic bias is considered significant when it is of the same order as the statistical uncertainty, quantitatively,

\begin{equation}
\frac{\Delta\Theta_i}{\sigma_{\Theta_i}} \sim \mathcal{O}(1).
\end{equation}

We use the methodology developed in \cite{Kumar:2025nwb} to account for the systematic errors in the waveform model $h(t;\mathbf{\Theta})$. In the frequency domain, the waveform model is denoted by $\tilde{h}(f;\mathbf{\Theta})$. The true model $\tilde{h}_\mathrm{true}(f;\mathbf{\Theta})$ may differ from the reference waveform $\tilde{h}_\mathrm{ref}(f;\mathbf{\Theta})$. Hereafter, we drop the explicit frequency and parameter dependence of waveform model strain and denote $\tilde{h}(f;\mathbf{\Theta})$ simply by $\tilde{h}$. The difference between the `true' waveform strain $\tilde{h}_\mathrm{true}$ and reference waveform $\tilde{h}_\mathrm{ref}$ can be parametrized in terms of amplitude and phase uncertainties,
\begin{eqnarray}
    \tilde{h}_\mathrm{true} &=& \tilde{A}_\mathrm{true}\exp\{i\tilde{\phi}_\mathrm{true}\} \nonumber\\
    &=& (\tilde{A}_\mathrm{ref} +\delta\tilde{A}_\mathrm{abs})\exp\{i( \tilde{\phi}_\mathrm{ref}+\delta\tilde{\phi}_\mathrm{abs})\}
\end{eqnarray}
\noindent
where $\tilde{\phi}_\mathrm{ref}$ is the phase evolution of the reference waveform model, $\delta \tilde{A}_\mathrm{abs}$ is the absolute amplitude error, $\delta\tilde{\phi}_\mathrm{abs}$ is the absolute error in phase in radians. The above equation can be rewritten in one of the following parametrizations \citep{Kumar:2025nwb},
\begin{eqnarray}
     \tilde{h}_\mathrm{true}&=& \tilde{h}_\mathrm{ref}(1+\delta\tilde{A})\exp\{i \tilde{\phi}_\mathrm{ref}\delta\tilde{\phi}_\mathrm{rel}\} \label{eqn:rel_phase}\\
     &=& \tilde{h}_\mathrm{ref}(1+\delta\tilde{A})\exp\{i \delta\tilde{\phi}_\mathrm{abs}\} \label{eqn:abs_phase},
\end{eqnarray}
\noindent
where $\delta \tilde{A}$ is the relative/fractional amplitude error, and $\delta\tilde{\phi}_\mathrm{rel}$ is the relative/fractional phase error. One may use either parametrizations represented by Eq. \eqref{eqn:abs_phase} (absolute phase errors) or Eq. \eqref{eqn:rel_phase} (relative phase errors) to modify the waveform model used in the standard \ac{pe} pipeline. This framework differs from the detector calibration correction scheme in two important respects: (i) the amplitude and phase corrections to the waveform approximants are applied in the signal frame, i.e.,  before projecting the waveform onto the detectors, and (ii) for current-generation detectors, waveform systematic uncertainties are expected to dominate over detector calibration uncertainties, therefore, the prior ranges assigned to waveform uncertainty parameters are broader than those used for the detector calibration parameters.

\subsection{Prior choices for $\delta \tilde{A}$ and $\delta\tilde{\phi}$}
The prior widths for the amplitude and phase uncertainties can be estimated from the expected contributions of waveform systematics, missing physical effects, and data analysis artifacts. Ideally, estimates of waveform-model uncertainties should be provided by the waveform-model developers or inferred from a detailed understanding of the waveform-modeling methodology. Similarly, uncertainties associated with missing physical effects may be estimated when the corresponding waveform signatures are sufficiently well understood. We assume the total error budget to be:
\begin{equation}
\Delta\Theta_\mathrm{total} = \Delta\Theta_\mathrm{WF} +\Delta\Theta_\mathrm{MP}+\Delta\Theta_\mathrm{DA} ,  
\end{equation}
\noindent
where $\Delta\Theta_\mathrm{WF}$ represents the error budget due to waveform systematics, $\Delta\Theta_\mathrm{MP}$ represents the error budget due to missing physical effects considered in the waveform model, and $\Delta\Theta_\mathrm{DA}$ represents the error budget due to any other data analysis artifacts. In general, the terms $\Delta\Theta_\mathrm{WF}$ and $\Delta\Theta_\mathrm{MP}$ can both be regarded as contributions to waveform systematics, though we distinguish between them in this work. The $\Delta\Theta_\mathrm{WF}$ term specifically represents the error budget specifically associated with waveform-modeling inaccuracies. 

When reliable estimates of these uncertainties are unavailable, one can remain agnostic, use sufficiently broad priors, and allow the data to constrain the uncertainty parameters. The intrinsic differences among waveform models may provide useful guidance and we may use them as proxies for the waveform uncertainties error budget. In this work, we take this approach, using sufficiently wide priors intended to account for the total error budget.

\subsection{Classification of waveform systematic errors}
In prior work \citep{Kumar:2025nwb}, we identified two classes of waveform systematic errors that can affect the \ac{pe} analysis. Type-A errors occur when the waveform model fails to accurately reproduce the true signal morphology, across the relevant parameter space. Type-B errors, on the other hand, arise when the waveform model can reproduce the signal accurately within the desired precision, but does so using incorrect parameters. The proposed method is most effective in mitigating type-A errors. If the waveform corresponding to the maximum-likelihood point in parameter space is sufficiently close to the true signal morphology (within a range that allows for corrections based on amplitude and phase uncertainties), we can expect to accurately reproduce the true signal morphology using the adjustments provided by these uncertainty parameters. For sufficiently loud signals, we can use the marginalized posterior distribution of the uncertainty parameters to determine whether type-A waveform systematic errors are present. In such cases, we expect waveform uncertainty parameters to exhibit posterior support away from zero.

Type-B errors are more challenging to address. If only one waveform model is affected by these errors, we can reduce the effect of these errors by using a mixture of posterior samples from the \ac{pe} runs of different waveform models. Additionally, we can construct a model-averaged likelihood function by utilizing multiple waveform models and assigning appropriate weights to the likelihoods associated with each model. These weights can be estimated based on how well a waveform model is calibrated to the nearest numerical relativity simulations within a given parameter space. In general, we expect that some combination of type-A and type-B errors might be present in a waveform model.  Mitigating waveform systematics in practice may therefore require combining multiple complementary approaches.

\section{Intrinsic differences between waveform models}
\label{sec:wf_differences}
In this section, we investigate the intrinsic differences between the waveform models in the parameter space spanned by the events considered in this study. From the posterior distributions of the events, we already know that different waveform models yield different  inference results from the same data. To complement this insight and isolate differences intrinsic to the waveform models themselves, we use the mismatch (to be defined later in this section) between the \ac{gw} waveform produced by these models for identical source parameters in the signal frame, i.e., before projecting them onto detectors. The closeness of the waveform template can be captured by the overlap between the two templates $h_1$ and $h_2$, which is defined as,
\begin{equation}
\mathcal{O}(h_1, h_2) = \frac{\langle h_1 |h_2 \rangle}{\sqrt{\langle h_1 | h_1 \rangle \langle h_2 | h_2 \rangle}}, \label{eqn:overlap}
\end{equation}
where the operation $\langle \cdot | \cdot \rangle$ represents the noise weighted inner product and is defined as,
\begin{equation}
    \langle h_1|h_2 \rangle = 4\Re\int_{f_\mathrm{min}}^{f_\mathrm{max}}\frac{\tilde{h}^*_1(f)\tilde{h}_2(f)}{S_{n}(f)} \; df. \label{eqn:inner_product}
\end{equation}
\noindent
Here, $\tilde{h}(f)$ represents the Fourier transform of the time domain strain $h(t)$, $h^*$ is the complex conjugate, and $S_n(f)$ represents the \ac{psd} of the noise.

The most direct comparison between two waveform models would be to calculate the overlap between two signals that differ only in the model used to produce them, with all other signal parameters held constant. However, significant differences could occur from (for our purposes) insignificant effects. For example, if two models were identical, apart from a small shift in the signal's coalescence time or an overall phase, we treat them as practically identical (i.e., mathematically equivalent). Any parameters whose differences we deem insignificant can be optimized over in the calculation of the inner product.

There is a considerable freedom in the choice of parameters to optimize over and therefore how to quantify waveform differences. As discussed in the appendix B of \citep{Schmidt:2014iyl}, properly optimizing over the signal's phase for precessing signals or signals with higher harmonics is nontrivial. Contemporary accuracy studies often optimize over time, coalescence phase and polarization angle \cite{Harry:2016ijz}. Additional optimizations are performed over specific spin angles \cite{Pratten:2020ceb}, and overlaps are sometimes averaged over the source's inclination \cite{Khan:2019kot}.  

For our illustrative purposes, we choose to quantify waveform differences with the overlap maximized over relative time and phase shifts between $h_1, h_2$, resulting in a quantity we refer to as the match $\hat{O}$. Correspondingly, the mismatch $\varepsilon$ is defined as
\begin{equation}
    \varepsilon = 1 - \hat{\mathcal{O}}(h_1, h_2) = 1 -  \max_{\phi_c,\, t_c} ~\mathcal{O}(h_1, h_2), \label{eqn:mismatch}
\end{equation}
\noindent
taking values between $0$ (equal waveforms) and $1$ (orthogonal waveforms).

\begin{figure*}
    \centering
    \includegraphics[scale=0.55]{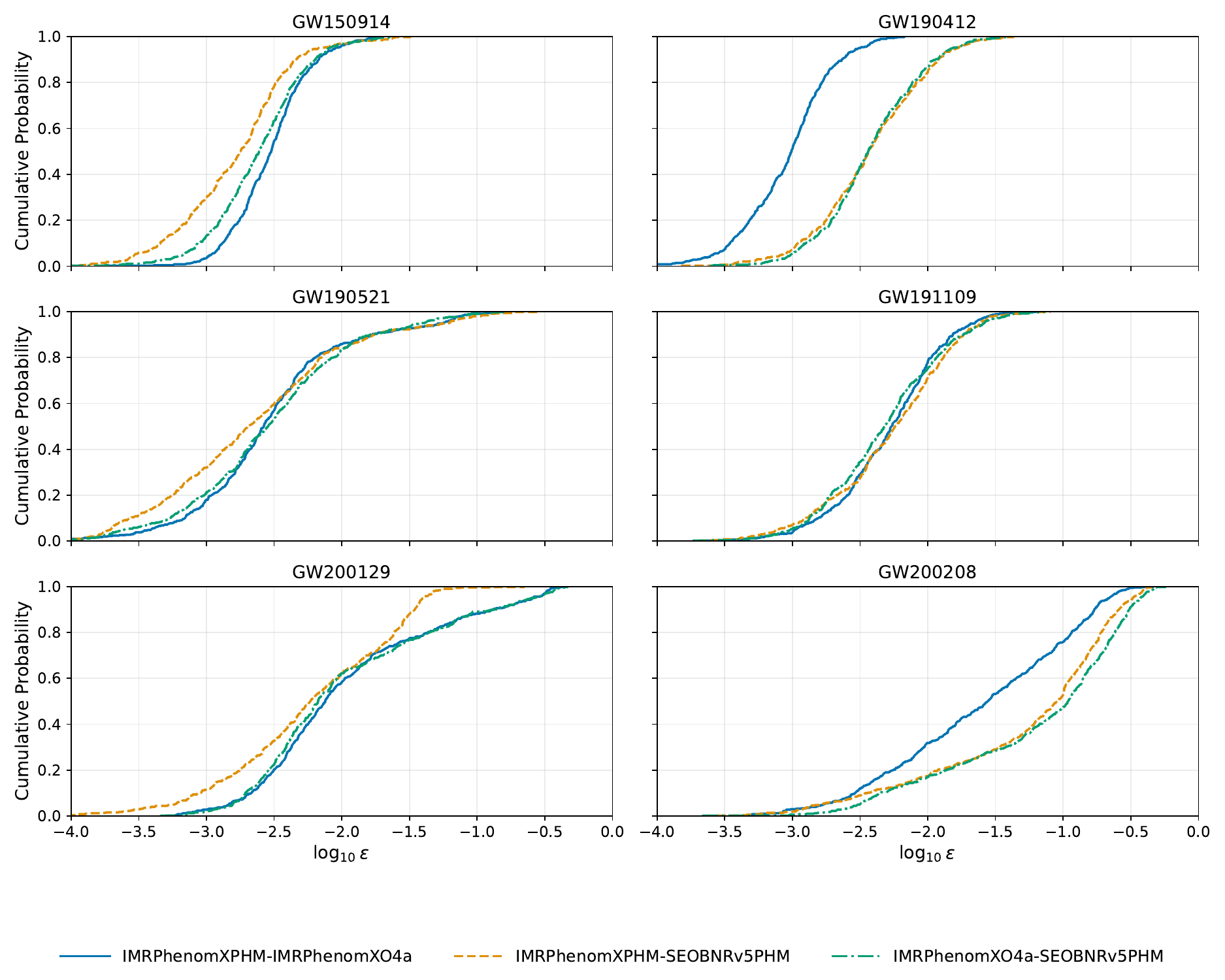}
    \caption{Here we show the cumulative distribution functions (CDF) of waveform mismatches for the \ac{gw} events considered in this study. Each panel shows the CDF of logarithm mismatch $\log_{10}\varepsilon$ for different waveform models of the $h_+$ polarization. The horizontal axis represents the logarithmic mismatch, while the vertical axis indicates the cumulative probability. Each curve represent mismatches between a pair of waveform models (see legends). We use frequency domain waveform models for this study.}
    \label{fig:mismatch_cdf}
\end{figure*}

For the waveform comparison, for a given event, we choose GWTC posterior samples \citep{gwtc21, gwtc3} corresponding to \ac{pe} run of one of the waveform models (\imrphenomxphm) and generate \ac{gw} strain using different waveform models. Here, we restrict the comparison to the intrinsic source parameters for the mismatch comparison. These intrinsic differences among the waveform models, quantified by the mismatches, limit the consistency of \ac{pe} runs with different waveform models. We randomly sample from the posterior distribution to generate multiple waveform strains based on the intrinsic parameters, such as masses and spins. Next, we estimate the match between these models. We use the frequency-domain implementations of following waveform models: \imrphenomxphm, \imrphenomxo, and \seobnr.  
We use the low-frequency cutoff of 20 Hz and the \texttt{aLIGOZeroDetHighPower} PSD model implemented in the \texttt{LALSimulation} \citep{lalsuite} package. Figure \ref{fig:mismatch_cdf} shows the cumulative distribution of mismatches for different pairs of waveform models from the posterior samples of different events. The results indicate that the similarity between the waveform models varies across different regions of parameter space corresponding to various events.

\begin{figure*}
    \centering
    \includegraphics[scale=0.7]{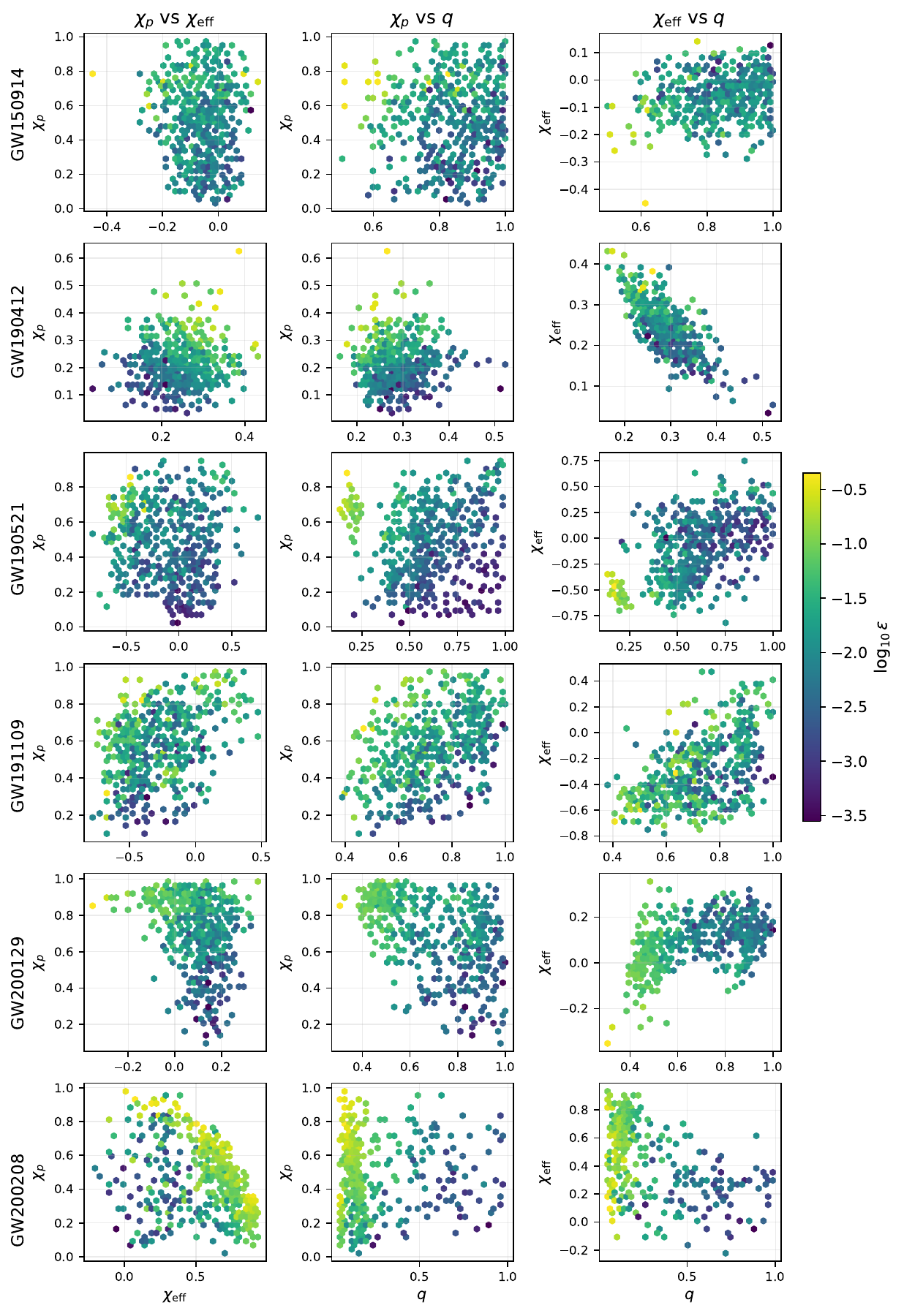}
    \caption{The hexbin plots showing the mean logarithmic waveform mismatch (between waveform pair \imrphenomxphm-\seobnr), $\log_{10}\varepsilon$, as a function of binary parameters for different gravitational-wave events considered in this study. Each row corresponds to one event, with the event name indicated on the left. Columns show: (1) $\chi_\mathrm{p}$ vs $\chi_{\rm eff}$, (2) $\chi_\mathrm{p}$ vs mass ratio $q$, and (3) $\chi_{\rm eff}$ vs $q$. Color represents the mean value of $\log_{10}\varepsilon$ in each hexagonal bin. The samples are drawn from the posterior probability distribution of each event with \imrphenomxphm waveform model. This figure highlights the region of parameter space where the intrinsic mismatch is higher. }
    \label{fig:scatter_mismatches}
\end{figure*}

In Figure \ref{fig:scatter_mismatches}, we identify the regions of parameter space with the largest mismatches for various events for the waveform model pairs: \imrphenomxphm and \seobnr. The highest mismatches occur in regions characterized by a high mass ratio \((q = {m_2}/{m_1} \leq 1)\), significant precession \((\chi_\mathrm{p})\), and large negative effective spin \((\chi_{\text{eff}})\). This trend is particularly evident when two of these parameters are explored jointly. The high-mass events GW191109\_010717 and GW200129\_065458 exhibit larger overall mismatches.

The accuracy of waveform models is known to vary across parameter space. Higher degrees of asymmetry in the system through unequal masses and complex spin dynamics make the signal structure richer and more difficult to model accurately. Numerical-relativity simulations are also more challenging and computationally demanding for high mass ratio systems or high spins. Therefore, current models are calibrated with fewer simulations in these regimes to achieve the same level of modeling accuracy that is achievable for non-spinning, near equal mass systems.

An effective mitigating strategy should account for this parameter dependence. In our framework, this means we need to incorporate priors on amplitude and phase uncertainties informed by our understanding of waveform uncertainties as a function of parameter space. An effective implementation of our framework  requires estimating the error budget $\Delta \Theta_\mathrm{WF} + \Delta\Theta_\mathrm{MP}$ and map it into amplitude and phase uncertainty priors as functions of frequency and parameter space. If $\mathbf{\Theta}$ are waveform parameters, then the distribution of amplitude and phase uncertainties is expected to depend on the waveform parameters, i.e., 

\begin{eqnarray}
\delta\phi (f_i, \mathbf{\Theta}) &\sim& \mathcal{N}_{i}(\mu=0, \sigma_{\delta \phi}(\mathbf{\Theta})) \\
\delta A (f_i, \mathbf{\Theta}) &\sim& \mathcal{N}_{i}(\mu=0, \sigma_{\delta A}(\mathbf{\Theta}))
\end{eqnarray}

\begin{table*}[]
\centering
\begin{ruledtabular}
\begin{tabular}{lll}
Sampling Parameters & Description & Prior \\ \hline

$\mathcal{M}^\mathrm{src}$ & Source-frame chirp mass 
& \multirow{2}{*}{Uniform in component masses} \\

$q = \frac{m_2}{m_1}~(m_1 \geq m_2)$ & Mass ratio & \\

$V_c$ & Comoving volume 
& Uniform in comoving volume  \\

$\iota$ & Inclination angle 
& $\pi(\iota) \propto \sin(\iota)$ \\

$a_i$ & Spin magnitude
& $\mathcal{U}(0, 0.99)$ \\

$\theta_i, \phi_i$ & Spin polar, azimuthal angle
& Uniform solid angle \\

$t_c$ & Time of arrival 
& Uniform: $\pi(t_c) = \mathcal{U}(t_c - 0.1,\, t_c + 0.1)$ \\

RA, dec & Right ascension and declination 
& Uniform sky \\

$\phi_c$ & Coalescence phase 
& Uniform angle: $\pi(\psi) = \mathcal{U}(0, 2\pi)$ \\

\hline

$\delta A (f_i)$ & Amplitude uncertainty parameter  
& $\mathcal{N}(\mu = 0,\, \sigma = 0.032)$\\ 

$\delta \phi_\mathrm{abs}(f_i)$ & Phase uncertainty parameters: absolute phase error 
& $\mathcal{N}(\mu = 0,\, \sigma = 0.14)$\\

$\delta \phi_\mathrm{rel}(f_i)$ & Phase uncertainty parameters: relative phase error
& $\mathcal{N}(\mu = 0,\, \sigma = 0.032)$\\

\end{tabular}
\end{ruledtabular}
\caption{This table describes the priors $\pi$ used in the \ac{pe} of simulated signals. The distribution $\mathcal{U}(a,b)$ refers to a one-dimensional uniform distribution defined within the interval $(a,b)$. For the WF-Error parameters, when we employ cubic spline curves with ten nodal points across the frequency range. The prior distributions for each $\delta\Tilde{A}_i$ and $\delta\Tilde{\phi}_i$ are specified accordingly. These sampling parameters are then transformed into the standard input parameters of waveform models (detector frame masses, luminosity distance, Cartesian spin components) using standard transformation functions.  }
\label{table:pe_priors}
\end{table*}

This notation indicates that both the phase uncertainty, $\delta\phi$, and the amplitude uncertainty, $\delta A$, follow normal distributions with standard deviations that depend on the waveform parameters. This way, we allow for larger priors for $(\delta A, \delta\phi)$ in the parameter region where the waveform model is more inaccurate and narrower priors where the waveform model appears to provide a good match with the true signal. In the present work, we do not employ the above recipe but use the same priors for uncertainty parameters across the parameter space.

\section{Applying the method on selected events from O1-O3}
\label{sec:events_analysis}
In this section, we describe \ac{pe} runs performed on the selected events from the first three observing runs, discussed in section \ref{sec:events_selection}. We use the waveform uncertainty parametrization with the framework developed in \citep{Kumar:2025nwb}. For each event, we conduct three sets of analyses:
\begin{enumerate}[label=(\roman*)]
    \item \textit{Baseline Run:} In this set of runs, we use a reference waveform model, $h_\mathrm{ref}$, and conduct the standard \ac{pe} analyses (without waveform uncertainty parameters). We refer to this set as the baseline runs.
    \item \textit{Absolute Phase Errors (APE):} In this analysis, we account for the amplitude and phase uncertainties in the waveform model $h_\mathrm{ref}$, with phase corrections parameterized by absolute phase errors as given by equation \eqref{eqn:abs_phase}. 
    \item \textit{Relative Phase Errors (RPE):} In this analysis, we account for the amplitude and phase uncertainties in the waveform model $h_\mathrm{ref}$, but we correct the phase using relative phase errors as shown in equation \eqref{eqn:rel_phase}.
\end{enumerate}
For each event, we use the Welch method to estimate the \ac{psd} from 512 seconds of data surrounding the event, using overlapping segments of 8 seconds each. By default, we keep the event at the centre of this 512-second window unless the data is contaminated or unavailable in this window. In that scenario, we shift the analysis window accordingly. We use the publicly available package \texttt{PyCBC Inference} \citep{Usman:2015kfa}, and the waveform systematics plugin \texttt{pycbc\_wferror\_plugin} \citep{GitHubPlugin} for parameter estimation. We use the publicly available nested sampler \textsc{Dynesty} \citep{speagle:2019} for Bayesian inference. In Table \ref{table:pe_priors}, we list the prior probability distribution used in the \ac{pe} runs.

Data from some events were processed using glitch-mitigation schemes. In these cases, deglitched frames were created by removing the glitches from the gravitational wave time-series data. We make use of these deglitched frame files whenever available. However, we also used the raw/unprocessed frame files downloaded from the \ac{gwosc} \footnote{\url{https://gwosc.org}}. Wherever applicable, we performed \ac{pe} runs on both raw and deglitched frame files. We will explicitly list these cases in the subsections below, which cover specific events. Even though our framework is designed primarily to mitigate the waveform systematic errors, we want to check if it can identify and mitigate certain types of systematic errors associated with the data analysis artifacts. 

For each scenario: event, type of run (baseline, APE, RPE), frame files (raw or deglitched), and waveform model, we perform 2-3 sets of runs based on computation resource availability. We set the random seeds for these runs to be different, and we consider the runs converged when different runs yield consistent posterior samples. Otherwise, we perform another set of identical runs with more stringent sampler settings, such as increasing the number of live points and the minimum number of walks before proposing a new live point. 

\subsection{Parameter estimation with O1-O3 events}
We perform three sets of analyses: i) baseline, ii) APE, and iii) RPE parametrization and compare the resulting posterior distributions with those reported in the GWTC \citep{gwtc21, gwtc3} and 4OGC catalogs \citep{Nitz:2021zwj}. For comparison, we use posterior samples from the \imrphenomxphm and \texttt{SEOBNRv4PHM} waveform models in the GWTC catalogs. The GWTC analyses make use of the publicly available \ac{pe} pipeline bilby \citep{bilby_paper}, while the 4OGC analysis uses the \texttt{PyCBC\_Inference} pipeline \citep{Biwer:2018osg}. There are multiple waveform models used in GWTC analysis, while 4OGC analysis employs the \imrphenomxphm waveform model.

Recently, a bug was identified in the treatment of detector calibration uncertainties in previous analyses \citep{Baka:2025bbb}. This error affected both the GWTC and 4OGC posterior samples. In the current analysis, we use a corrected implementation of the calibration uncertainty framework. Further details are provided in appendix \ref{appendix:calibration_correction}. Next, we discuss each event considered in this work in temporal order.

\subsubsection{GW150914\_095045}
\begin{figure*}
    \centering
    \includegraphics[scale=0.58]{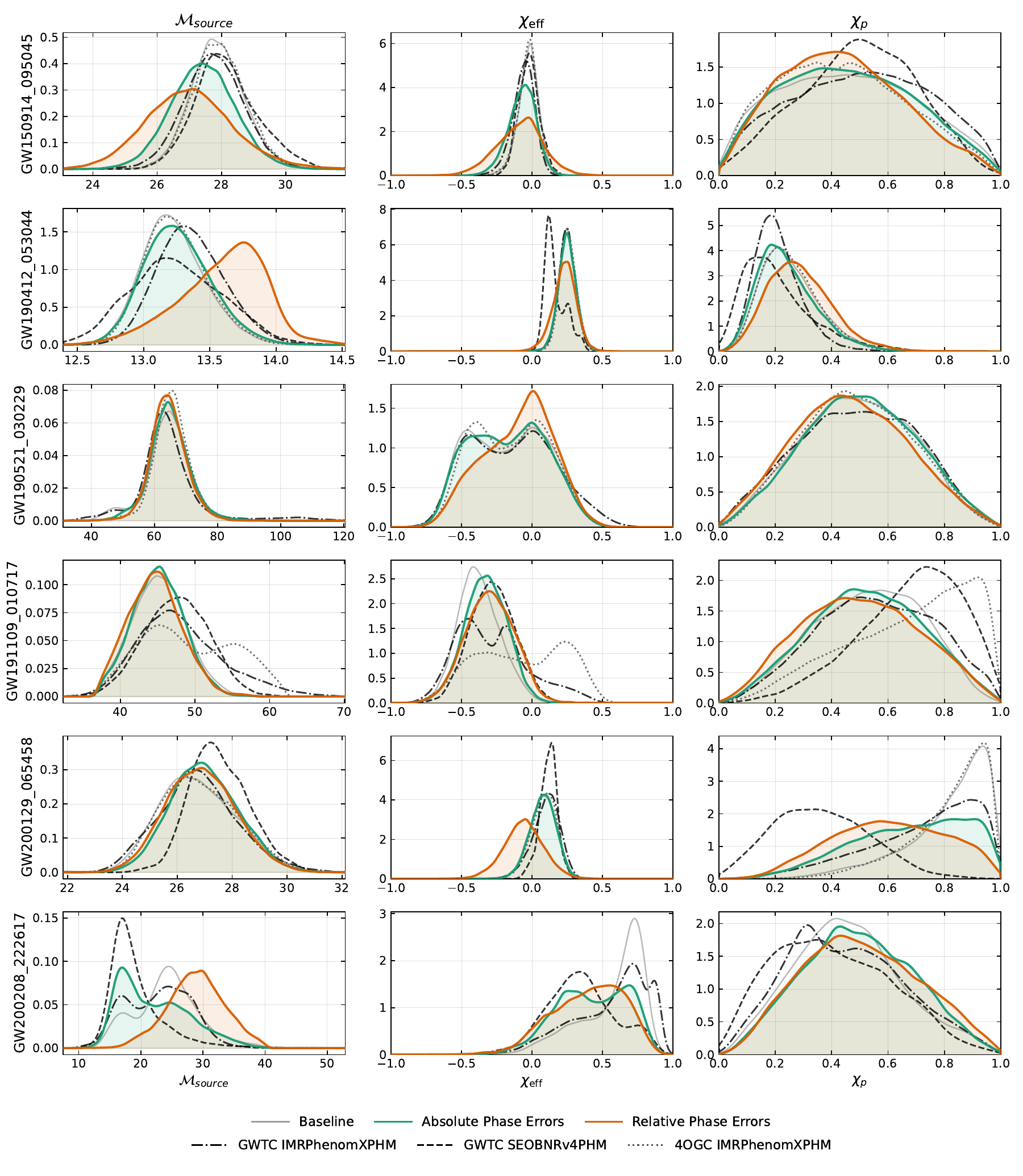}
\caption{This figure summarizes the \ac{pe} result of different events for three key parameters: source frame chirp mass ($\mathcal{M}_\textrm{source}$), precessing spin parameter ($\chi_{p}$), and effective spin parameter ($\chi_\mathrm{eff}$). The solid Grey curves represent the baseline runs (with the \imrphenomxphm waveform model), while the solid green and orange curves represent the absolute-phase waveform-error and relative-phase waveform-error runs, respectively. For reference, we also show the equivalent baseline runs from the GWTC and OGC catalogs. Shaded regions highlights the 1D marginalized posteriors for the \ac{pe} runs with waveform uncertainties. Modulo GWTC and 4OGC catalogs, all other runs for this figure make use of raw frame files from \ac{gwosc}.}
    \label{fig:summary_plot}
\end{figure*}

Even though the GW150914\_095045 runs did not show any signs of systematic errors across studies, we still use it as an example of how the framework behaves when no significant systematic errors are present. The mismatch study (Figures \ref{fig:mismatch_cdf} and \ref{fig:scatter_mismatches}) indicate that this event shows the minimal intrinsic difference across different waveform models. We observe that (see figure \ref{fig:summary_plot}) there are no significant differences between the baseline runs and runs with waveform uncertainties. There is a slight shift in the median value of the source-frame chirp mass and a widening of the posterior samples, as expected. For this event, there are no deglitched files, so all \ac{pe} runs are done using the \ac{gwosc} raw frame files. We get a slight increase ($\approx 0.4\%$) in the maximum value of \ac{snr} for the absolute phase uncertainty run as well as the relative phase uncertainty run. However, the difference in the natural logarithm of Bayesian evidence favours the Baseline model over the relative phase errors. In table \ref{tab:log_evidence_snr_ratios}, we list the difference in log Bayes evidence between the waveform uncertainty parametrization and the baseline model.   

\subsubsection{GW190412\_065458}
The event GW190412\_065458 is one of the events with an asymmetric mass ratio and a positive effective spin parameter $\chi_\mathrm{eff}$ \citep{LIGOScientific:2020stg}. In GWTC-2.1 analysis, the reported values of the effective spin parameter are $\chi_\mathrm{eff}=0.25^{+0.10}_{-0.10}$ (\imrphenomxphm) and $0.14^{+0.17}_{-0.07}$ (\texttt{SEOBNRv4PHM}). The 4OGC analysis recovered $\chi_\mathrm{eff}=0.25^{+0.10}_{-0.09}$ (\imrphenomxphm). The recovered mass ratio parameter ($q$) for GWTC-2.1 analysis is $q=0.28^{+0.09}_{-0.07}$ (\imrphenomxphm), $0.41^{+0.12}_{-0.15}$ (\texttt{SEOBNRv4PHM}), and for 4-OGC analysis is $0.27^{+0.08}_{-0.07}$ for \imrphenomxphm waveform model. Due to its highly asymmetric mass ratio and non-zero effective spin in a $90\%$ confidence interval, this event is considered in various studies to be coming from formation channels such as repeated black hole mergers in a dense cluster \citep{Rodriguez:2020viw}, or as one of the black holes as a remnant of a previous black hole merger \citep{Gerosa:2020bjb}.

\begin{table*}[t]
\centering
\begin{ruledtabular}
\begin{tabular}{lp{1pt}ccp{1pt}cc}
\toprule
Event && $\left\langle \ln\left(\frac{E_{\mathrm{APE}}}{E_{\mathrm{baseline}}}\right) \right\rangle$ & $\left\langle \ln\left(\frac{E_{\mathrm{RPE}}}{E_{\mathrm{baseline}}}\right) \right\rangle$ & & $\frac{\delta \rho_{\mathrm{abs}}}{\rho_{\mathrm{baseline}}}\times 100~(\%)$ & $\frac{\delta \rho_{\mathrm{rel}}}{\rho_{\mathrm{baseline}}}\times 100~(\%)$ \\
\hline
\midrule
GW150914\_095045 && $-0.74$ & $-6.98$ && $0.40$ & $0.47$ \\
GW190412\_053044 && $-0.26$ & $-8.61$ && $0.55$ & $2.58$ \\
GW190521\_030229 && $0.33$ & $0.84$ && $1.31$ & $0.44$ \\
GW191109\_010717 && $1.85$ ($2.42$) & $2.43$ ($3.16$) && $0.97$ ($1.09$) & $1.59$ ($1.20$) \\
GW200129\_065458 && $2.51$ ($0.16$) & $-0.74$ ($-3.55$) && $0.21$ ($0.41$) & $0.29$ ($1.00$) \\
GW200208\_222617 && $1.04$ & $0.09$ && $-0.57$ & $0.85$ \\
\bottomrule
\end{tabular}
\end{ruledtabular}
\caption{Comparison of log-evidence ratios and fractional \ac{snr} increase for absolute-phase (APE) and relative-phase errors (RPE) parametrizations with respect to the baseline configuration using the waveform model \imrphenomxphm. The angular brackets $\langle ...\rangle $ represent the average of multiple runs done in each scenario. Additionally, we show the percentage change in maximum value of \ac{snr}: the quantities $\delta \rho_\mathrm{abs}$ and $\delta \rho_\mathrm{rel}$ represent the increase in maximum \ac{snr} when using the APE and RPE waveform uncertainty parametrizations, respectively, compared to the baseline \ac{pe} \ac{snr}.  All quantities are estimated from the raw frame files from \ac{gwosc}, except for those in brackets (...), which are based on deglitched frame files.}
\label{tab:log_evidence_snr_ratios}
\end{table*}

\begin{figure*}
    \centering
    \includegraphics[scale=0.8]{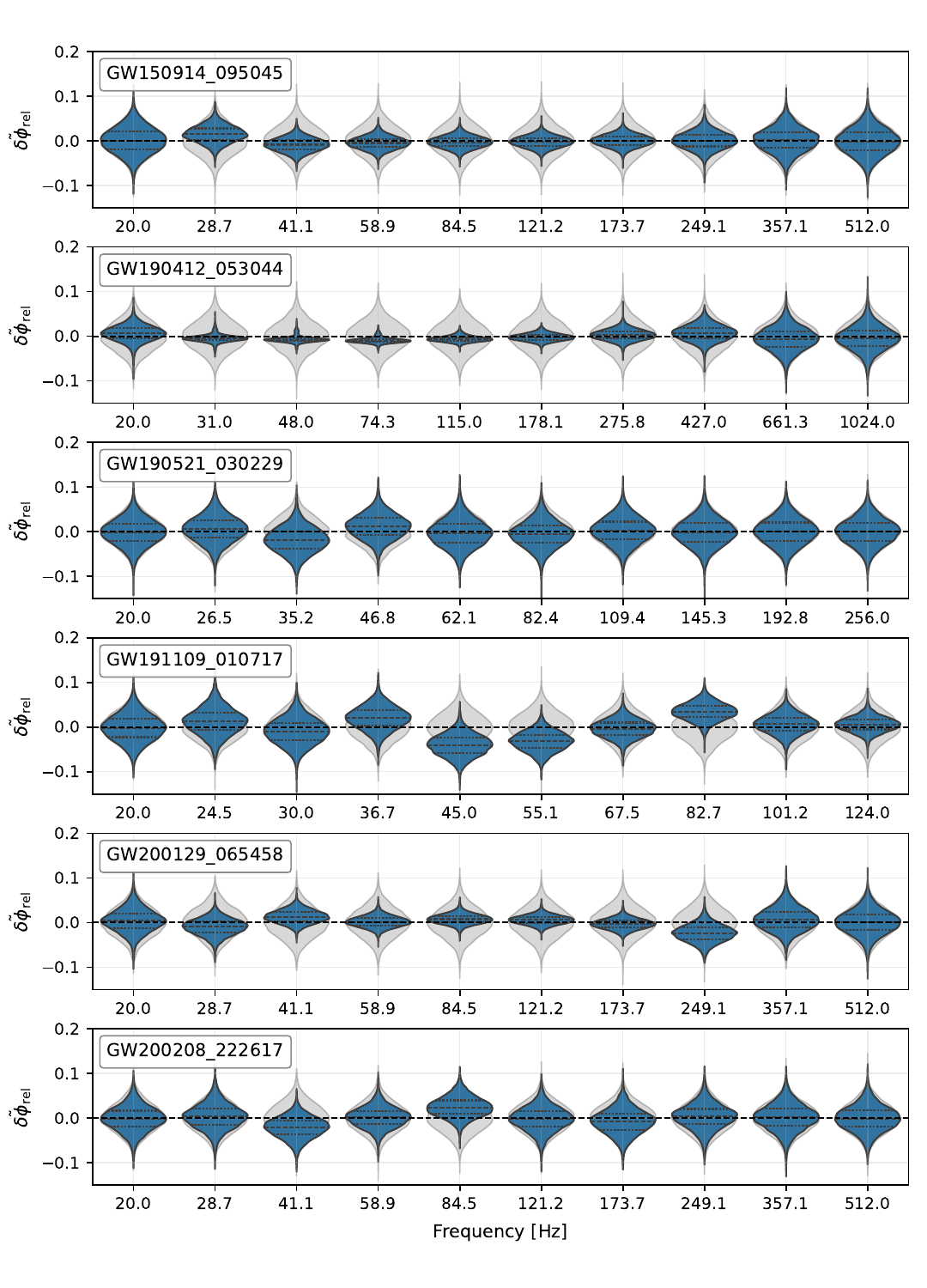}
    \caption{The recovery of relative phase uncertainties, $\tilde{\delta\phi}_\mathrm{rel}(f)$, at the nodal points. The light-grey shaded regions represent the prior, while the dark-blue shaded region indicates the posterior recovery. We use \imrphenomxphm~waveform model for these runs. The dashed vertical line represent the `no-modification' line to the baseline waveform model. Posteriors converging away from the dashed line represent presence of type-A waveform systematic errors. The representation is chosen such that the width of the probability distribution is same.}
    \label{fig:dphi_imrphenomxphm}
\end{figure*}

In our analysis, we recover the effective spin parameter to be $0.25^{+0.10}_{-0.10}$ for baseline \ac{pe} run with \imrphenomxphm waveform. For waveform uncertainty parametrizations, we recover $\chi_\mathrm{eff}=0.25^{+0.11}_{-0.10}$ (APE) and $0.23^{+0.13}_{-0.15}$ (RPE) with same waveform model. Both the parametrizations still support the higher value of $\chi_\mathrm{eff}$ with $90\%$ confidence interval above $\chi_\mathrm{eff}=0$. We also observe a higher value of the source-frame chirp mass in the RPE parametrization recovery (Figure \ref{fig:summary_plot}). We also observe that the phase uncertainty parameters at different nodal points in the RPE model are most constrained in this event for interior frequency nodal points (see Figure \ref{fig:dphi_imrphenomxphm}). We observe a $\sim0.5\%$ increase in maximum \ac{snr} for APE parametrization compared to the baseline \ac{pe} run, while the same for RPE parametrization is $\sim2.6\%$. The Bayesian evidence for the baseline model is significantly stronger than that for the RPE parametrization ($\sim8.61$ in favour of baseline parametrization), while the APE parametrization and the baseline model show similar evidence.
\subsubsection{GW190521\_065458}
The event GW190521\_065458 \citep{LIGOScientific:2020iuh} is a high mass signal where the significant probability distribution of the primary black hole mass lies in the mass gap produced by the pair-instability supernova process. The final mass of the remanent black hole is an intermediate mass black hole (IMBH) \citep{LIGOScientific:2020ufj}. This event could have formed through various formation channels, such as hierarchical mergers of smaller black holes in dense environments, such as star clusters or active galactic nuclei \citep{LIGOScientific:2020ufj, Morton:2023wxg}. In the GWTC-2.1 analysis, the \imrphenomxphm model shows a small bimodality in the source-frame chirp mass (Figure \ref{fig:summary_plot}). The spin parameter $\chi_\mathrm{eff}$ also shows bimodality with one peak at negative $\chi_\mathrm{eff}$ and another near zero for both GWTC-2.1 and 4OGC analysis.

In our framework, for the baseline \ac{pe} run, we observe a small bimodality at the lower end of the chirp mass; however, this bimodality disappears for the APE and RPE parametrizations. For the APE parametrization, we still observe a bimodal $\chi_\mathrm{eff}$ distribution, whereas for the RPE model, the effective spin parameter favours the second peak. For the phase-uncertainty parameters, this event does not provide stronger constraints relative to the priors used. The recovered parameters for waveform uncertainty parametrization do not show significant differences from the baseline model, i.e., $\delta\tilde{\phi}_\mathrm{rel}=0$. We do observe a $\sim1.3\% (0.44\%)$ increase in maximum \ac{snr} for APE (RPE) parametrization relative to baseline \ac{pe} runs, while no such increase is observed for RPE parametrization. The Bayesian evidence remains similar across all three \ac{pe} runs.

\subsubsection{GW191109\_010717} 
The event GW191109\_010717 is particularly significant because it exhibits anti-aligned spin properties, which may hint at a dynamical origin \citep{Zhang:2023fpp}. However, this event was affected by a scattered light glitch that occurred very close to the signal in the LIGO Livingston detector. In the GWTC-3 analysis, both waveform models, \texttt{SEOBNRv4PHM} and \imrphenomxphm, show multimodality and support for negative $\chi_\mathrm{eff}$: $-0.31^{+0.53}_{-0.32}$ and $-0.28^{+0.26}_{-0.26}$ for \imrphenomxphm and \texttt{SEOBNRv4PHM} respectively. In the 4-OGC analysis, the $\chi_\mathrm{eff}$ parameter shows two modes, with greater support for the negative mode. 

Subsequent analyses of this event have revealed indications of anti-aligned spins \citep{Udall:2024ovp}, spin precession, and even hints of eccentricity \citep{Romero-Shaw:2022xko, Gupte:2024jfe}. A detailed study conducted by \citep{Udall:2024ovp} employs two different models for the glitch and performs a combined inference of glitch and the signal. Their findings indicate that when a flexible glitch model—based on sine-gaussian wavelet reconstruction—is used, a bimodal distribution of effective spins $\chi_\mathrm{eff}$ emerges, reflecting both aligned and anti-aligned spins. Conversely, when a scattered light glitch model is applied, the parameter estimation favors the anti-aligned spin distribution. 

In our study, we perform \ac{pe} runs with both deglitched and raw frame files. We also make use of three waveform models: \imrphenomxphm, \imrphenomxo, and \nrsur. We find that when we conduct baseline \ac{pe} runs, the marginalized posterior distributions of specific parameters differ across waveform models and among the frame files (see Figure \ref{fig:summary_plot}). When we include waveform uncertainties in the \ac{pe}, we find that both raw and deglitched frame files yield comparatively consistent results. Moreover, the differences between the waveform models are also reduced. The recovered distribution of the $\chi_\mathrm{eff}$ across all waveform models parameter favors an anti-aligned spin configuration. For the \imrphenomxphm waveform model, and raw frame files, we find the recovered values of $\chi_\mathrm{eff}=-0.38^{+0.28}_{-0.22}$ (Baseline), $-0.33^{+0.24}_{-0.25}$ (APE), and $-0.30^{+0.28}_{-0.28}$ (RPE). For deglitched frames, the inferred values of $\chi_\mathrm{eff}=-0.18^{+0.52}_{-0.39}$ (Baseline), $-0.21^{+0.47}_{-0.33}$ (APE), and $-0.24^{+0.36}_{-0.37}$ (RPE). In our analysis of the \imrphenomxphm waveform model, we found that the average difference in the natural log-evidence for this event is 1.85 (2.42) in favor of the APE parametrization compared to the baseline \ac{pe} run when using raw (deglitched) frame files. The same results apply for the RPE parametrization, with an average difference of 2.43 (3.16) when using the same type of frame files. This indicates that, despite a significantly larger number of additional waveform-uncertainty parameters, the Bayesian evidence favors both the APE and RPE parametrizations over the baseline waveform model.

Additionally, we observe that at the frequency nodal points beyond 35 Hz, the relative phase errors of the waveform model exhibit a slight deviation from zero, although this could still be consistent with a Gaussian noise realization. 

\begin{figure*}
    \centering
    \includegraphics[scale=0.85]{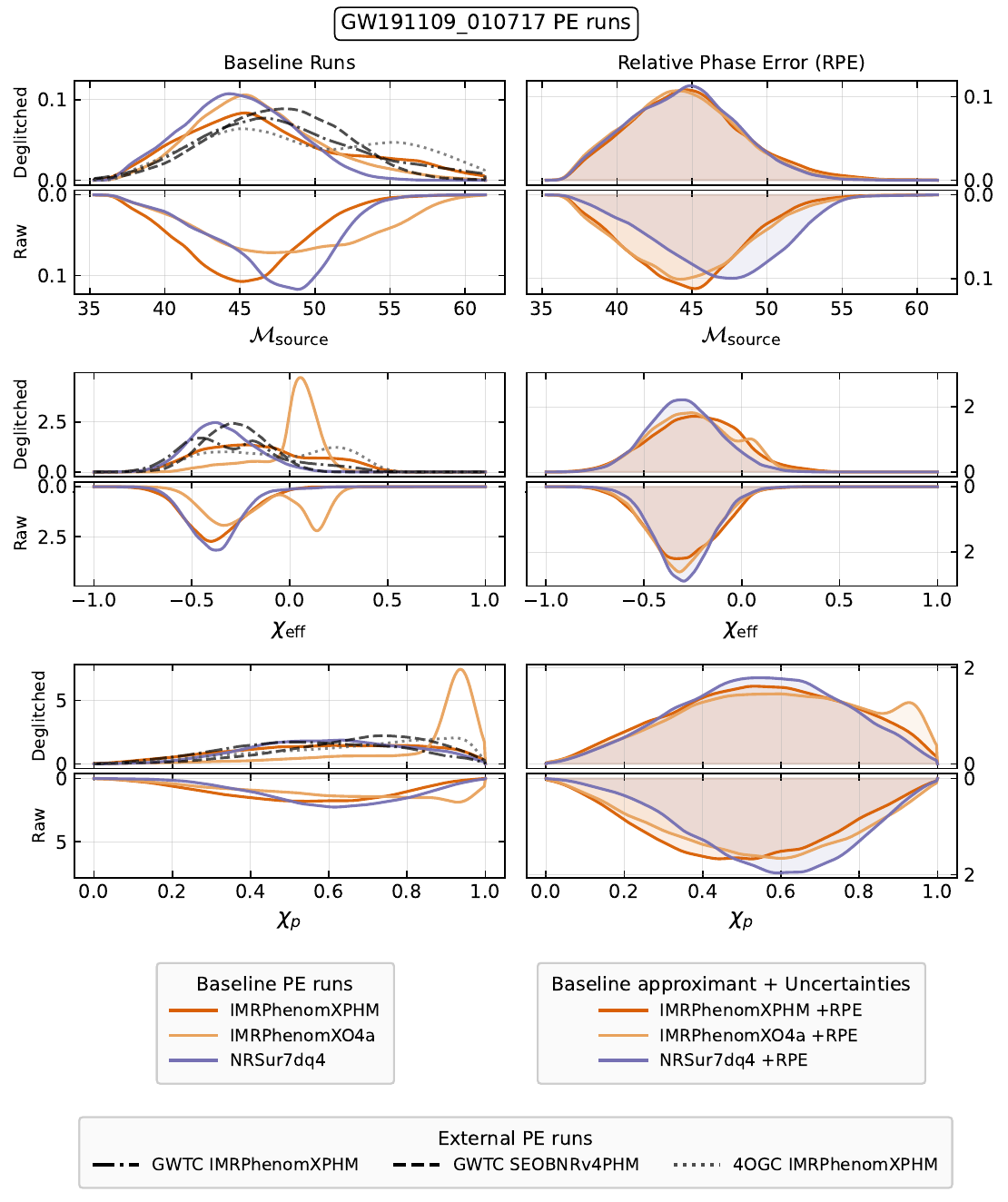}
    \caption{This figure illustrates the differences in 1D marginalized posterior samples for event GW191109{\_}010717, comparing two sets of frame files: deglitched and raw. The deglitched frames are those in which a glitch mitigation scheme is applied, while no such mitigation scheme is applied in raw frames. The first column represents the baseline \ac{pe} runs that do not include waveform uncertainties, while the second column shows \ac{pe} results from models that incorporate uncertainties, specifically relative phase error. The first, second and third rows shows 1D marginalized posteriors for source frame chirp mass ($\mathcal{M}_\mathrm{source}$), spin-precession ($\chi_\mathrm{p}$), and effective spin parameter ($\chi_\mathrm{eff}$) respectively. In each row, the top panel displays results from the runs using deglitched frame files, whereas the inverted plots represent the results from the raw frame files. We use three different waveform models in these runs: \imrphenomxphm, \imrphenomxo, and \nrsur. The plots that incorporate waveform uncertainties are shaded for clarity. For reference, we also show the GWTC and 4OGC \ac{pe} runs in left panels.}
    \label{fig:raw_vs_deglitched_gw191109}
\end{figure*}

Figure \ref{fig:raw_vs_deglitched_gw191109} shows the difference in the 1D marginalized posterior samples between different waveform models (\imrphenomxphm, \imrphenomxo, and \nrsur) for the parameters $M_\mathrm{source}, \chi_\mathrm{eff}$, and $\chi_p$ when baseline \ac{pe} run. We also observe differences between the \ac{pe} runs in the raw and deglitched frame files for a given waveform model. However, when we use the waveform uncertainty parametrization, the \ac{pe} runs give comparatively consistent results regardless of the waveform model or the type of frame file used. It is more pronounced in RPE parametrization compared to APE parametrization (not shown in the figure).

\begin{figure*}
    \centering
    \includegraphics[scale=0.8]{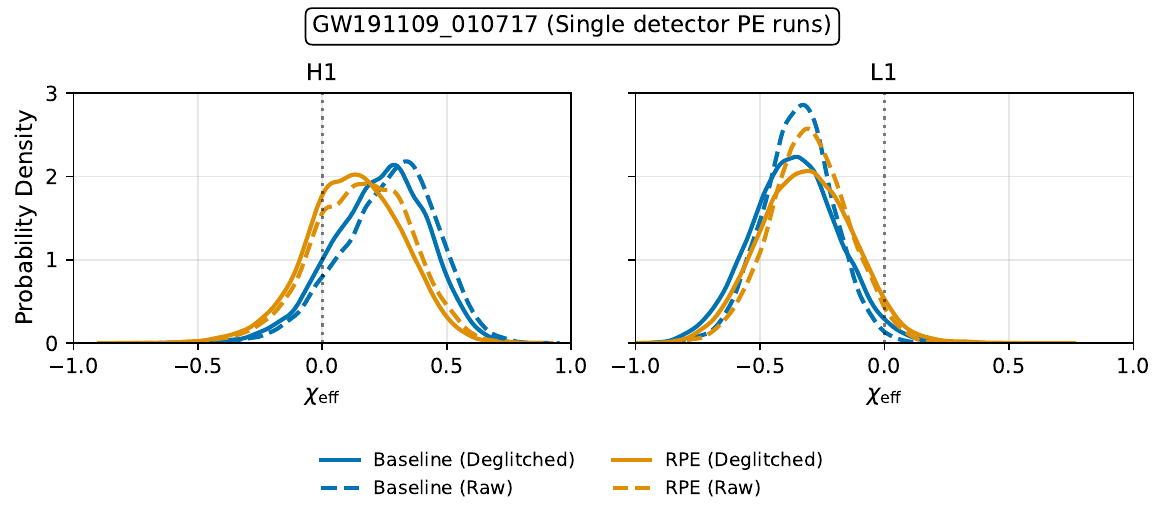}
\caption{The figure shows the 1D marginalized posterior distribution for the single detector runs for the event GW191109. We use \imrphenomxphm~ waveform model for these runs. The solid curves correspond to the \ac{pe} runs with deglitched frame files, while dashed curves represent the same for raw \ac{gwosc} frame files. The vertical dotted line serves as a reference point for $\chi_\mathrm{eff}=0$. The left panel shows the results from the LIGO-Hanford detector, while the right panel shows the results from the LIGO-Livingston detector. It can be seen that the support for $\chi_\mathrm{eff} < 0$ comes from the LIGO-Livingston detector, consistent with the observation of \citep{Udall:2024ovp}.}
    \label{fig:gw191109_single_detector_chi_eff}
\end{figure*}

\begin{figure*}
    \centering
    \includegraphics[scale=0.8]{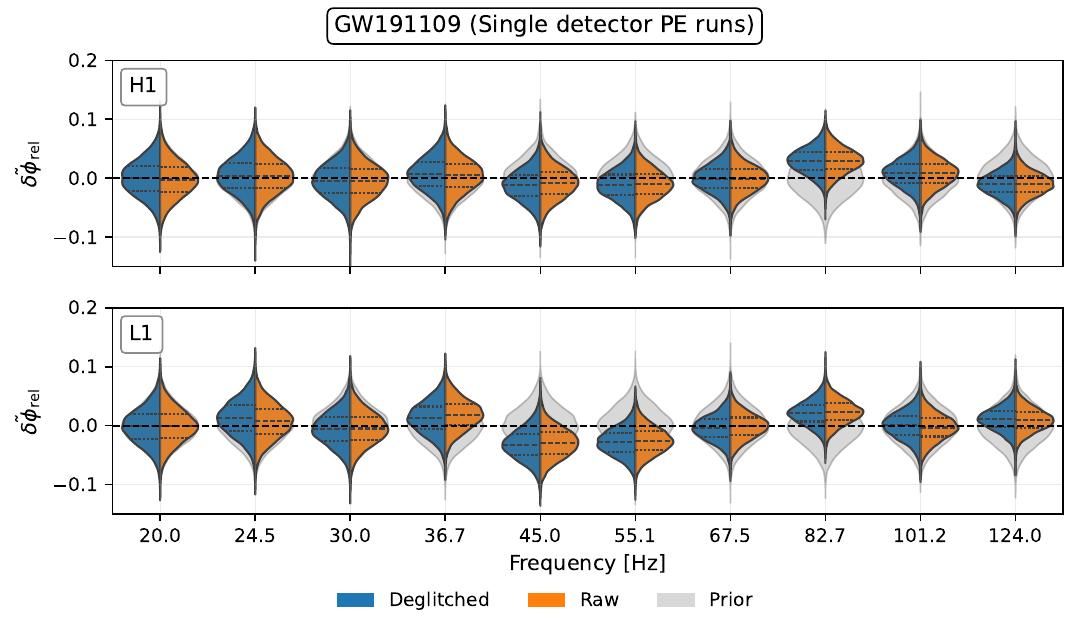}
\caption{The figure shows the 1D marginalized posterior samples of the waveform uncertainty parameter $\delta\tilde{\phi}_\mathrm{rel}$ for various frequency nodal points chosen for the \ac{pe} run with RPE parametrization for single-detector \ac{pe} runs. The top panel shows the results from the LIGO-Hanford detector, while the bottom panel shows the results from the LIGO-Livingston detector. The blue half-violin plots represent the \ac{pe} runs from deglitched frame files, while the orange half-violin plots represent the \ac{pe} results from raw \ac{gwosc} frame files. The grey-shaded region represents the prior distribution for each nodal point. There was a glitch present in the LIGO-Livingston data at around $\sim40$ Hz or $0.1-0.04$ seconds before the merger \citep{Udall:2024ovp}. The deglitched frame files are constructed by modelling and removing the best-fit glitch model in the LIGO-Livingston detector.}
    \label{fig:gw191109_single_detector_dphi}
\end{figure*}

We also perform a set of single-detector runs (with \imrphenomxphm~waveform model) for this event and find that the negative $\chi_\mathrm{eff}$ support comes from the Livingston detector. The posterior samples obtained using data from the Hanford detectors have a single mode at $\chi_\mathrm{eff}>0$. These observations are consistent among baseline runs and RPE parametrization runs. In a previous study \citep{Udall:2024ovp}, the authors noted that the negative effective spin measurement is attributed to the segment of the LIGO Livingston data between 30 and 40 Hz, or 0.1-0.004 seconds before the merger, which coincides with the excess glitch power. In figure \ref{fig:gw191109_single_detector_chi_eff}, we show 1D marginalized posterior samples for the $\chi_\mathrm{eff}$ parameter. For the \ac{pe} runs using deglitched frame files, the maximum value of \ac{snr} for the LIGO-Hanford detector is 9.7 (9.8) for baseline (RPE) run, and for the LIGO-Livingston detector is 12.8 (13.1) for baseline (RPE) \ac{pe} runs. In the case of the raw frame files, the maximum  \ac{snr} for the LIGO-Hanford detector is 9.6 (9.8) for baseline (RPE) run, and for the LIGO-Livingston detector is 13.5 (13.7) for baseline (RPE) \ac{pe} runs. We did not find any significant difference in the log evidence between baseline runs and RPE parametrization runs for the single-detector \ac{pe} runs. In figure \ref{fig:gw191109_single_detector_dphi}, we show the recovery of the waveform uncertainty parameter $\delta\tilde{\phi}_\mathrm{rel}$ for RPE parametrization for single detector \ac{pe} runs. It can be seen that the deviation pattern noted in figure \ref{fig:dphi_imrphenomxphm} for GW191109\_010717 is similar to that obtained for a single-detector run using the LIGO-Livingston data.

To quantify the effect of the presence of a glitch on the signal, we also perform simulations in which a GW191109-like signal is injected into Gaussian, stationary noise, with and without a glitch in one of the detectors. We will discuss the results of this simulation study in a subsequent sub-section (\ref{subsec:simulations}).

\subsubsection{GW200129\_065458}
GW200129\_065458 is another interesting event that may have originated through a dynamical formation channel, especially considering its reported high precession. The GWTC-3 catalog \citep{gwtc3} reported potential waveform systematic issues with this event. The LIGO Livingston data were affected by quality issues, and glitch-mitigation techniques were used to remove a glitch near the signal. In GWTC-3 analysis, the \imrphenomxphm model constrains the precessing spin parameter $\chi_\mathrm{p} = 0.77^{+0.19}_{-0.44}$ while \texttt{SEOBNRv4PHM} waveform model infer it to be $\chi_\mathrm{p} = 0.36^{+0.31}_{-0.25}$. In 4-OGC analysis, with \imrphenomxphm model, $\chi_\mathrm{p}$ was inferred to be $\chi_\mathrm{p} = 0.85^{+0.11}_{-0.32}$.

This event has been analyzed by subsequent studies in support of the claim of precession. In \citep{Hannam:2021pit}, the authors investigated the event using the \nrsur waveform model and found that the binary's orbit precesses at a rate ten orders of magnitude faster than the previous weak-field measurement of binary pulsars. They also found a very high spin for the primary black hole in the binary. Another study used a neural network-based approach to mitigate broadband noise in the LIGO Livingston detector and found that evidence of precession remains \citep{Macas:2023wiw}. Some studies constrain the kick velocity of this event to be $\geq 698 $km/s at the $90\%$ credible interval \citep{Varma:2022pld}.
\begin{figure*}
    \centering
    \includegraphics[scale=0.85]{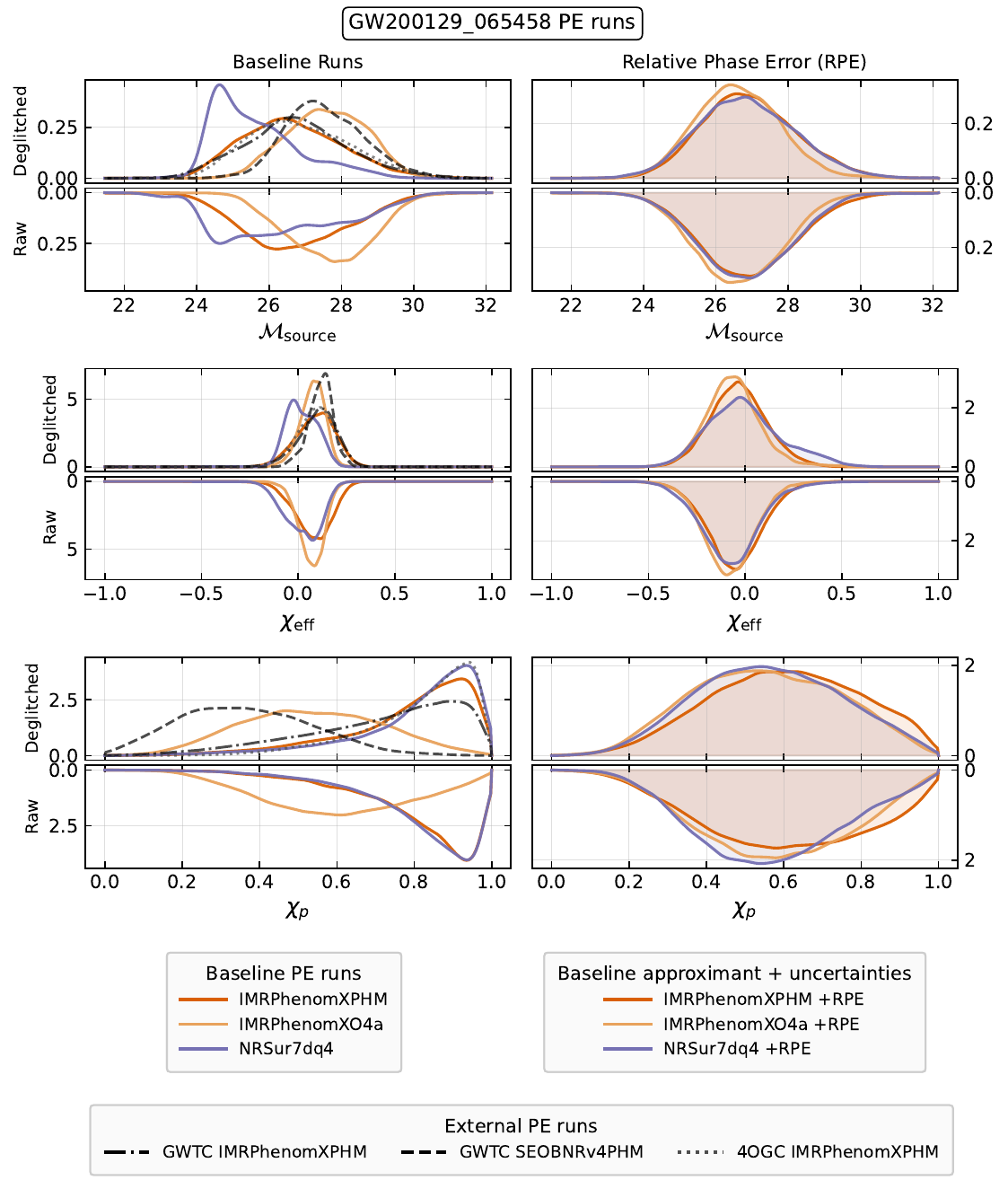}
    \caption{This figure illustrates the differences in 1D marginalized posterior samples for event GW200129{\_}065458, comparing two sets of frame files: deglitched and raw. The deglitched frames are those in which a glitch mitigation scheme is applied, while no such mitigation scheme is applied in raw frames. The first column represents the baseline \ac{pe} runs that do not include waveform uncertainties, while the second column shows \ac{pe} results from models that incorporate uncertainties, specifically relative phase error. The first, second and third rows shows 1D marginalized posteriors for source frame chirp mass ($\mathcal{M}_\mathrm{source}$), spin-precession ($\chi_\mathrm{p}$), and effective spin parameter ($\chi_\mathrm{eff}$) respectively. In each row, the top panel displays results from the runs using deglitched frame files, whereas the inverted plots represent the results from the raw frame files. We use three different waveform models in these runs: \imrphenomxphm, \imrphenomxo, and \nrsur. The plots that incorporate waveform uncertainties are shaded for clarity. For reference, we also show the GWTC and 4OGC \ac{pe} runs in left panels.}
    \label{fig:raw_vs_deglitched_gw200129}
\end{figure*}

For this event, we conducted parameter estimation runs on both the raw frame files and the deglitched frames. Our findings support the presence of high-spin precession within the waveform uncertainty framework; however, the magnitude of the recovered spin precession is lower than that observed in the baseline runs. When doing \ac{pe} with raw frame files, we constrain the spin precession parameter to be $\chi_\mathrm{p}=0.85^{+0.12}_{-0.36}$ (baseline), $\chi_\mathrm{p}=0.70^{+0.25}_{-0.38}$ (APE), and  $\chi_\mathrm{p}=0.60^{+0.31}_{-0.33}$ (RPE). With deglitched frame files, we find the constraints on $\chi_\mathrm{p} = 0.83^{+0.14}_{-0.39}$ (Baseline), $\chi_\mathrm{p} = 0.67^{+0.27}_{-0.38}$ (APE), and $\chi_\mathrm{p}=0.60^{+0.31}_{-0.32}$ (RPE). With waveform uncertainties, the APE parametrization yields a larger $ \chi_\mathrm{eff}$ than the RPE uncertainty parametrization (see Figure \ref{fig:summary_plot}). Additionally, we note that the phase uncertainty parameters for this event are more tightly constrained than for several other events, such as GW190521\_065458, GW191109\_010717, and GW200208\_222617, due to high \ac{snr}. For the raw frame files, we notice $0.2\% (0.3\%)$ increase in \ac{snr} for APE (RPE) parametrization compared to the baseline \ac{pe} runs. For the deglitched frame files, we notice $\sim0.4\%~(1\%)$  increase in maximum value of \ac{snr} for APE (RPE) parametrization compared to the baseline \ac{pe} run. The APE parametrization has stronger Bayesian evidence compared to the baseline model for both raw and deglitched frame files, while the baseline model is preferred over the RPE parametrization (see table \ref{tab:log_evidence_snr_ratios}).

Figure \ref{fig:raw_vs_deglitched_gw200129} demonstrates the differences in 1D marginalized posterior samples for selected parameters ($\mathcal{M}_\mathrm{source}, \chi_\mathrm{eff},$ and $\chi_p$) for different waveform models used in \ac{pe} for this event. It also highlights variations in the posterior distributions across different frame files for the same waveform model. When the waveform mitigation framework (RPE parametrization) is applied, these discrepancies, both between waveform models and across frame files, are significantly reduced.

\subsubsection{GW200208\_222617}
GW200208\_222617 is a comparatively low \ac{snr} (network \ac{snr} $\sim 8$) event that shows multimodal mass posterior \citep{gwtc3}. It also shows inconsistency in the $\chi_\mathrm{eff}$ parameter between \imrphenomxphm~ ($\chi_\mathrm{eff} = 0.62 ^{+0.26}_{-0.59}$) and \texttt{SEOBNRv4PHM} ($\chi_\mathrm{eff} = 0.34 ^{+0.45}_{-0.38}$) waveform models, with large positive $\chi_\mathrm{eff}$ for both models. The authors in \citep{Romero-Shaw:2025vbc} explored this event for potential signatures of eccentricity. Notably, this event does not appear in the 4-OGC catalog. Around the event, there were no data quality concerns or glitches therefore the runs are performed on raw frame files.

We report that we still recover the positive $\chi_\mathrm{eff}$ value in the waveform uncertainty framework. Our baseline \ac{pe} runs gives the value of $\chi_\mathrm{eff}=0.65^{+0.17}_{-0.56}$ while APE parametrization run yields  $\chi_\mathrm{eff}=0.43^{+0.35}_{-0.42}$, and RPE parametrization runs give $\chi_\mathrm{eff}=0.42^{+0.33}_{-0.46}$. The APE model run shows multimodality in the effective spin parameter and captures both modes from the GWTC-3 analysis: \texttt{SEOBNRv4PHM} (lower mode) and \imrphenomxphm~ (higher mode). On the other hand, RPE parametrization runs do not show multimodality in $\chi_\mathrm{eff}$, and it peaks roughly between the two modes. Given that this is a low SNR event, the constraints on phase uncertainty parameters are weak (see Figure \ref{fig:dphi_imrphenomxphm}). The APE model shows the decrease in the maximum \ac{snr} value ($-0.6\%$) compared to the baseline model but it could be due to sampling errors due to low \ac{snr}. The RPE parametrization shows the increase in maximum \ac{snr} by $0.85\%$ compared to the baseline \ac{pe} run. The Bayesian evidence for RPE and baseline model is indistinguishable while APE model is slightly preferred over baseline model (natural log evidence ratio $\sim1$ in favor of APE parametrization).
\begin{figure*}
    \centering
    \includegraphics[scale=0.8]{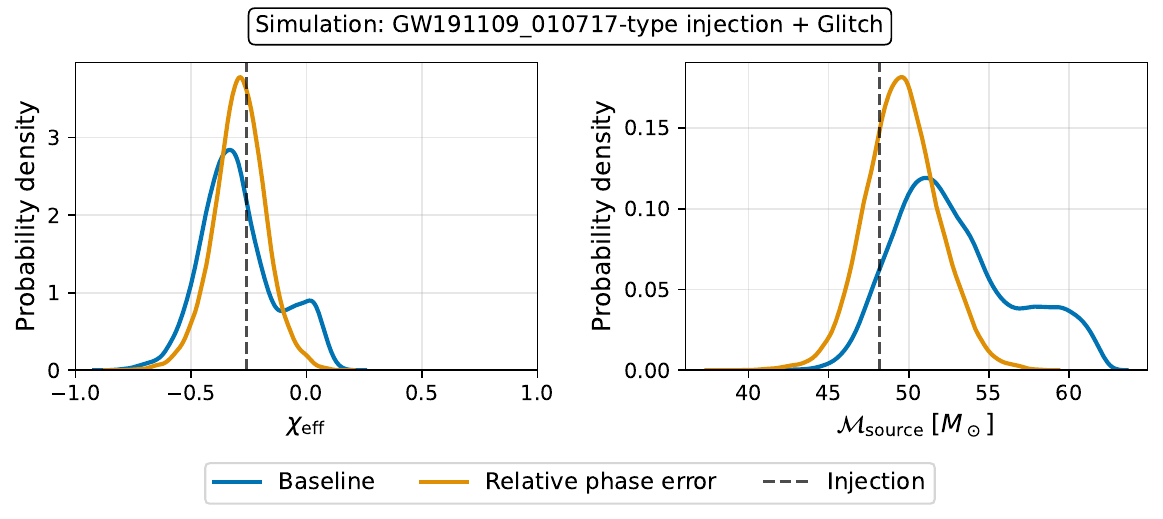}
\caption{Here we show the 1D marginalized posterior samples for parameters: $\chi_\mathrm{eff}$ (left) and the source frame chirp mass $\mathcal{M}_\mathrm{source}$, for a simulated GW191109-type system with a scatter light glitch in the vicinity of the signal in the LIGO-Livingston detector. The blue curve represents the baseline \ac{pe} run without waveform uncertainties, while the orange curve represents a \ac{pe} run that includes waveform uncertainty parameters. The baseline run exhibits bimodality due to the presence of the glitch in one detector, whereas the run with the RPE parameterization does not show this bimodality. The dashed vertical line represents the injected value. We use the \imrphenomxphm waveform model for injection and recovery for these \ac{pe} runs.}
    \label{fig:gw191109_simulations_chi_eff_mchirp}
\end{figure*}

\subsection{Simulations}
\label{subsec:simulations}
To study the effectiveness of waveform-uncertainty parameterization in mitigating systematic effects caused by glitches, we use a GW191109-type system, inject it into Gaussian noise, and generate two frame files for the LIGO Hanford and LIGO-Livingston detectors. We then create another simulated frame file for the LIGO-Livingston detector by injecting a scatter light glitch near the signal. We use the whitened strain for a scatter light glitch, taken from the \texttt{GravitySPY} database \citep{Zevin:2016qwy}. The duration of the scatter light glitch is 2 seconds, and we inject it so that it overlaps with the signal: it starts $\approx1.5$ seconds before the merger and lasts $\approx0.5$ seconds after the merger.  

\begin{figure*}
    \centering
    \includegraphics[scale=0.8]{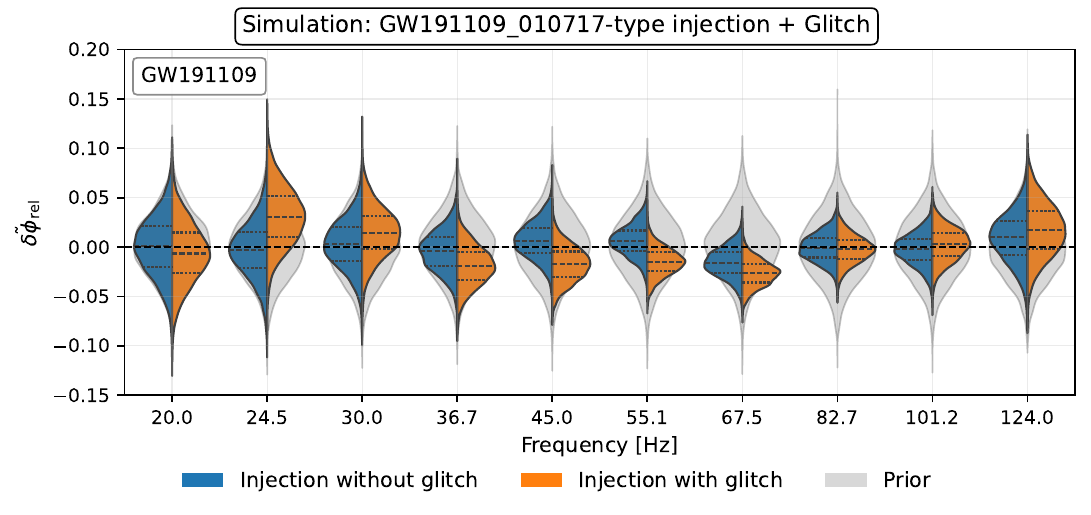}
\caption{The figure illustrates the recovery of the relative phase uncertainty parameter $\delta\tilde{\phi}_\mathrm{rel}$ for multiple nodal points. These \ac{pe} runs are done with a simulated GW191109-like signal in Gaussian noise in a two-detector network mimicking LIGO-Hanford and LIGO-Livingston detectors (blue). In the second set of simulations, we also inject a scatter-light glitch into the LIGO-Livingston detector near the signal (see the text for details). The \ac{pe} runs corresponding to the second set are shown in orange. The light-grey shaded region indicates the prior probability distribution. In this example, the presence of glitch produces larger deviation in $\delta\tilde{\phi}_\mathrm{rel}$ away from zero.}
    \label{fig:gw191109_simulations_dphi_rel}
\end{figure*}

We perform baseline and waveform-uncertainty parametrization \ac{pe} runs on both sets of frame files: with and without glitch. When we perform \ac{pe} runs on the frame files containing a glitch in the LIGO-Livingston detector, we observe bimodality in the $\chi_\mathrm{eff}$ and the source-frame chirp mass, $\mathcal{M}_\mathrm{source}$. This bimodality disappears when we perform \ac{pe} using waveform-uncertainty parametrizations. No such bimodality is observed in the \ac{pe} runs using frame files, which do not contain an injected glitch. In figure \ref{fig:gw191109_simulations_chi_eff_mchirp}, we show the recovery of the $\chi_\mathrm{eff}$ and $\mathcal{M}_\mathrm{source}$ for simulated data with a glitch in the LIGO-Livingston detector. We observe bimodality in the recovery of both parameters for the baseline \ac{pe} run. The bimodality disappears for \ac{pe} runs with RPE parametrization. In figure \ref{fig:gw191109_simulations_dphi_rel} we show the recovery of phase uncertainty parameter $\delta\tilde{\phi}_\mathrm{rel}$ for RPE parametrization. We notice that in the presence of a glitch, $\delta\tilde{\phi}_\mathrm{rel}$ deviates more from the reference line of no modification $\delta\tilde{\phi}_\mathrm{rel}=0$.

We want to highlight that a more detailed study, including an injection and recovery campaign with different types of glitches and signals, is needed to quantify the effectiveness of this method in the presence of glitches. Although this method is not specifically intended to address systematic errors in the presence of glitches, we found it beneficial for mitigating these errors in both our simulations and real data. The glitch mitigation/subtraction schemes may leave some residual power from the glitches and may affect parameter estimation, especially the inference of spins \citep{Udall:2025bts}.  We find that the waveform uncertainty parametrization can be useful in such scenarios and this needs to be investigated further. 

\section{Discussion and Summary}
\label{sec:summary}
The systematic errors in Bayesian inference for \ac{gw} merger events become particularly significant for high-\ac{snr} events where they can be comparable to the statistical uncertainties. Several observed \ac{gw} events already exhibit systematic biases that cannot be neglected. In a previous work, we developed a framework to address the systematic waveform errors by incorporating amplitude and phase uncertainty parameters in the waveform model. If we know the waveform modelling uncertainties in a given region of parameter space (e.g., provided by waveform developers), we can use them as prior distributions on these uncertainty parameters to mitigate potential systematic errors. In the absence of prior knowledge of waveform uncertainties, we can be agnostic and use sufficiently broad priors to capture potential systematic errors. 

In this work, we take selected events from the first three observing runs of the  LVK collaboration and apply our mitigation framework. We use two waveform uncertainty parametrizations corresponding to absolute and relative phase errors. Some events, such as GW191109\_010717 and GW200129\_065458, had data quality issues, and deglitched data frame files are provided for these events. We perform parameter inference on raw (non-deglitched) as well as deglitched frame files. For GW191109\_010717, the baseline runs are inconsistent across two types of frame files, even for the same waveform model. However, when we apply our mitigation framework, the runs show consistency between the frame files. We emphasize that even though the framework is not designed to perform \ac{pe} in the presence of glitches or to mitigate the effects of the deglitching process. Nevertheless, we find that the framework improves consistency between analyses performed on raw and deglitched data for the glitch-affected events GW191109\_010717 and GW200129\_065458.

One of the key features of this mitigation framework is its ability to determine if there are significant differences between the best-fit waveform model and the underlying signal. We can examine the posterior samples of the recovered amplitude and phase uncertainty parameters. If these values begin to converge away from zero (which serves as the reference point for no modifications to the baseline waveform model), it suggests the presence of systematic errors, and the waveform model is attempting to find the correct fit by accommodating the additional flexibility provided by these uncertainty parameters. In this study, we did not find significant deviations from zero in the uncertainty parameters for the events analyzed. However, there are indications of deviation in the phase uncertainty parameter of GW191109\_010717, particularly near the glitch frequency of 36 Hz.

To further investigate the origin of these effects, we performed additional single-detector \ac{pe} runs, and injection-recovery studies for GW191109\_010717 with and without a scatter-light glitch. We report that the major contribution for $\chi_\mathrm{eff}<0$ comes from the LIGO-Livingston data. Furthermore, we simulated a scenario in which a GW191109-like signal was injected into Gaussian noise and a scatter-light glitch in one of the detectors. We find bimodality in $\chi_\mathrm{eff}$ and in the source-frame chirp mass in baseline runs. This bimodality disappears when we use waveform uncertainty parametrization.

\subsection{Bayesian model comparison and interpretation of results}
Interpreting the results of the \ac{pe} analyses with uncertainty parameters requires care when comparing them with standard \ac{pe} runs. In this data-driven framework, the additional uncertainty parameters increase the prior volume and therefore introduce an Occam's razor penalty \citep{Trotta:2005ar} in the Bayesian evidence relative to the baseline model. Hence, the model with uncertainty parameters may not necessarily be favored by the evidence-based comparisons. A lower evidence value for the APE/RPE model compared to the baseline model does not indicate the absence of unresolved systematic effects. However, a larger evidence value, despite the increased model complexity suggests that the data favor the flexibility introduced by the uncertainty parametrization. When the Bayesian evidence is not decisive, we interpret the results conditional on the adopted waveform model and the assumed amplitude and phase uncertainty priors. 

In the absence of detailed knowledge of the error budgets for different types of systematics, we assume that the priors on the uncertainty parameters incorporate combined error budget. We then use the posterior distributions of the amplitude and phase uncertainty parameters as diagnostics for potential systematic effects. A posterior distribution significantly away from zero may indicate the presence of type-A waveform systematics or contamination from a nearby glitch. Exploring the frequency dependence of these recovered uncertainty parameters can also help identify regions where the baseline waveform model fails to accurately describe the signal.

\subsection{Limitations and Future Directions}
While we demonstrate the effectiveness of mitigating systematic errors in simulated and real events using \ac{gw} waveform uncertainty parametrization in a data-driven approach, we would like to address certain caveats and assumptions now. The main issue in this framework is the choice of the prior. In the data-driven approach, we assume prior distributions for the amplitude and phase uncertainty parameters and let the signal's loudness and the accuracy of the waveform model determine the constraints on these uncertainty parameters. If we choose priors that are too small, we might not be able to fully correct for the systematics. On the other hand, overly broad priors may weaken the constraints on the other physical parameter. A more principled approach would be to estimate error budgets. One way to do this is to provide probabilistic waveform models or uncertainty envelopes alongside waveform models. 

Another potential improvement that we intend to implement in the near future is using parameter-space-dependent priors on uncertainty parameters. For a typical system, we use wide astrophysical priors on masses and spin parameters. We have seen in Figure \ref{fig:scatter_mismatches} that the mismatch between waveform models becomes large for more asymmetric mass ratios, large precessing values, and high effective spin parameters. If we assume that the intrinsic differences among the waveform models are a reasonable proxy for the differences with the `true' signal, then we need to adopt comparatively wider uncertainty priors in that parameter space. If we know the inaccuracies of waveform models across parameter space, we can then translate them into amplitude and phase uncertainty bands and use them as prior distributions in \ac{pe} runs. 

Another limitation of the current framework may be realized when the cubic-spline method fails to capture the functional form of the true deviation from the underlying `correct' waveform model. It can happen when either the number of nodal points is insufficient or the functional form of the deviation is too complicated to be captured by a single blanket grid. To address the first point, we present a recipe in our method paper \citep{Kumar:2025nwb} that investigates how many frequency nodes are sufficient to capture a particular form of deviation we are looking for. For the second scenario, before the \ac{pe} run, we can estimate the functional form of uncertainty parameters as a function of frequency across the prior range, and then adjust the cubic-spline construction accordingly. 

Another important issue to be addressed in the future version of this analysis is to disentangle the systematic errors arising from waveform modeling uncertainties, missing physical effects, and data analysis artifacts, such as the presence of a glitch or shortcomings in glitch mitigation schemes. Understanding the error budget for each of these factors could help us clarify and separate the overlapping influences. To estimate the error budgets for the waveform-model uncertainties and missing physical effects, we can utilize the Fisher matrix formalism along with simulations. For the data analysis artifacts, we need to conduct a comprehensive injection-recovery campaign that covers various types of glitches and signals across the parameter space.

The proposed framework for doing \ac{pe} has a direct impact on similar analyses that rely on small modifications of waveform models, such as testing the general theory of relativity, or looking for micro-lensing signatures. In the current framework, the best way forward would be to estimate the waveform uncertainty error budget and develop statistical tools that indicate when the sought-after effects exceed the systematic errors.

\acknowledgments
We acknowledge support by the research programme of the Netherlands Organisation for Scientific Research~(NWO), the Max Planck Gesellschaft and the Max Planck Independent Research Group Program, through which this work was supported. We thank the computing team from AEI Hannover for their significant technical support. We thank Tomeszk Baka, and Mick Wright for fruitful discussions and comments. We also thank Lucy Thomas and Pierre Mourier for critical inputs on the manuscript. We also thank GRASP gravitational wave group members at Utrecht University and Binary Merger Observations and \ac{nr} group members at AEI Hannover for their feedback and valuable comments. We used the Holodeck computing cluster at AEI Hannover for all the computation. This work make use of open source software packages NumPy \citep{harris2020array}, SciPy \citep{2020SciPy-NMeth}, Matplotlib \citep{Hunter:2007}, PyCBC \citep{Usman:2015kfa}, Dynesty \citep{speagle:2019}, and LALSuite \citep{lalsuite}. We acknowledge the use of free version of ChatGPT (OpenAI) and Claude Code (Anthropic) in assisting with the review and refinement of plotting scripts used in this work.  All suggestions were evaluated and verified by the authors, who take full responsibility for the final results.

This research has made use of data or software obtained from the Gravitational Wave Open Science Center (gwosc.org), a service of the LIGO Scientific Collaboration, the Virgo Collaboration, and KAGRA. This material is based upon work supported by NSF's LIGO Laboratory which is a major facility fully funded by the National Science Foundation, as well as the Science and Technology Facilities Council (STFC) of the United Kingdom, the Max-Planck-Society (MPS), and the State of Niedersachsen/Germany for support of the construction of Advanced LIGO and construction and operation of the GEO600 detector. Additional support for Advanced LIGO was provided by the Australian Research Council. Virgo is funded, through the European Gravitational Observatory (EGO), by the French Centre National de Recherche Scientifique (CNRS), the Italian Istituto Nazionale di Fisica Nucleare (INFN) and the Dutch Nikhef, with contributions by institutions from Belgium, Germany, Greece, Hungary, Ireland, Japan, Monaco, Poland, Portugal, Spain. KAGRA is supported by Ministry of Education, Culture, Sports, Science and Technology (MEXT), Japan Society for the Promotion of Science (JSPS) in Japan; National Research Foundation (NRF) and Ministry of Science and ICT (MSIT) in Korea; Academia Sinica (AS) and National Science and Technology Council (NSTC) in Taiwan.

\section*{Data Release}
The data that support the findings of this article are
openly available \citep{GitHubPlugin, kumar_wferror_o1o2o3}, embargo periods may apply.

\appendix
\section{Additional details of \ac{pe} runs}
\label{appendix:additional_pe_details}

\begin{figure*}
    \centering
    \includegraphics[scale=0.58]{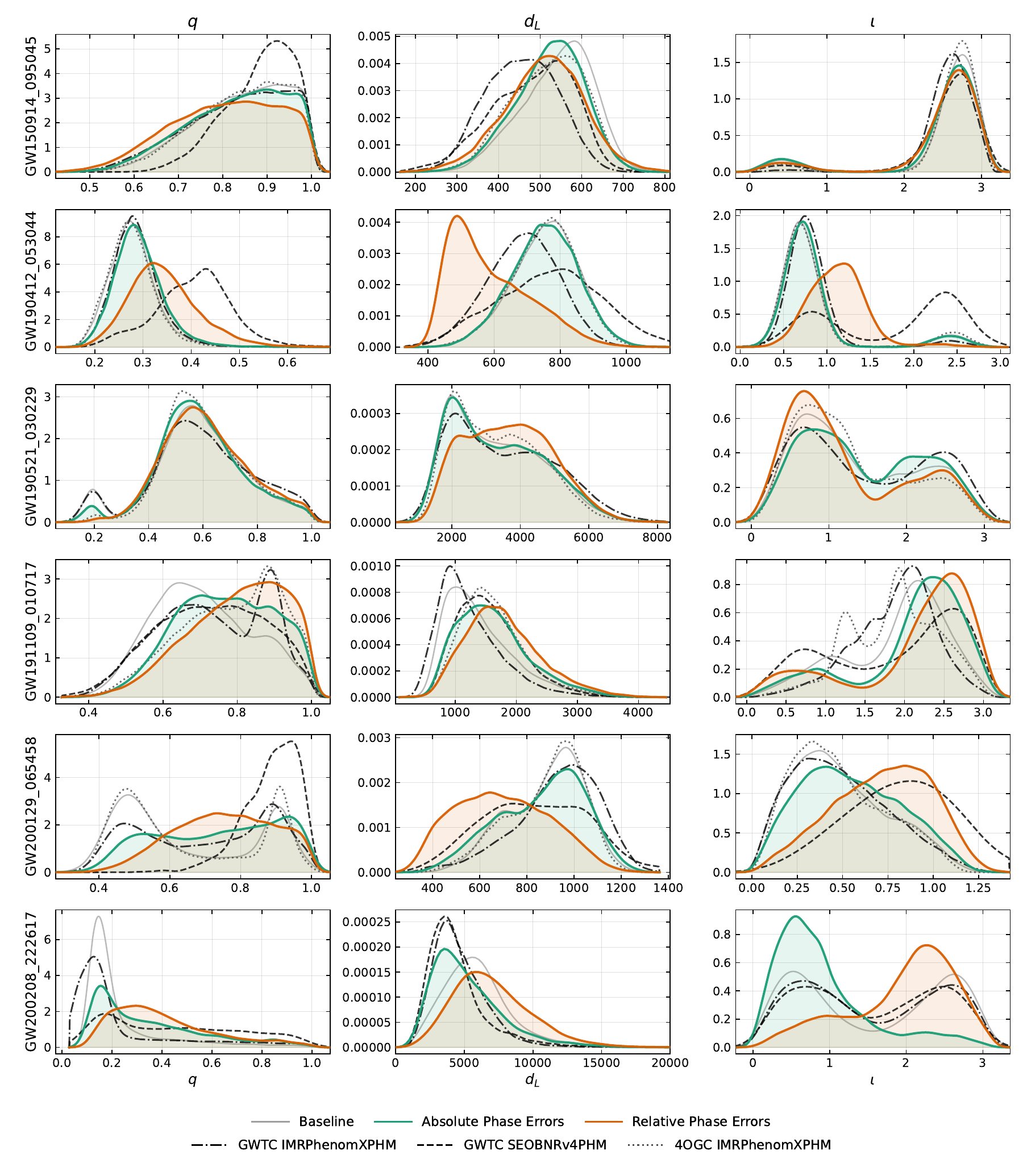}
\caption{This figure summarizes the \ac{pe} result for different events with the following parameters: mass ratio ($q=m_2/m_1$), luminosity distance ($d_L$), and inclination angle ($\iota$). The solid grey curves represent the baseline runs (with the \imrphenomxphm~waveform model), while the solid green and orange curves represent the absolute-phase waveform-error and relative-phase waveform-error runs, respectively. For reference, we also show the equivalent baseline runs from the GWTC and OGC catalogs. Shaded regions highlight the 1D marginalized posteriors for the \ac{pe} runs with waveform uncertainties. All the runs for this figure make use of raw frame files from \ac{gwosc}.}
    \label{fig:summary_plot_additional}
\end{figure*}

Here, we provide additional details regarding the \ac{pe} runs that are not covered in the main body of the paper. In figure \ref{fig:summary_plot}, we covered the parameters $\mathcal{M}_\mathrm{source}$, $\chi_\mathrm{eff}$, and $\chi_{p}$. In figure \ref{fig:summary_plot_additional}, we show the recovery of mass ratio $(q=m_2/m_1)$, luminosity distance ($D_L$), and inclination angle ($\iota$) for the events considered in this study. We use raw \ac{gwosc} frame files for these \ac{pe} runs. 

\begin{figure*}
    \centering
    \includegraphics[scale=0.58]{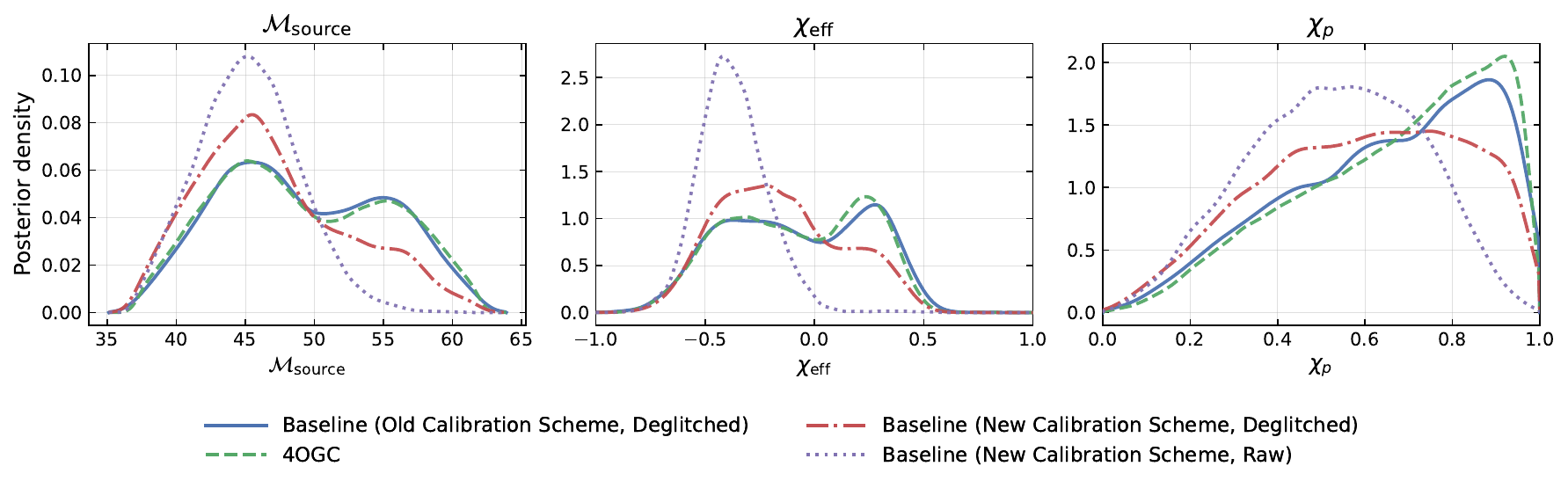}
\caption{This figure highlights the differences in the \ac{pe} runs for the old and new calibration schemes for the event GW191109. The 4OGC runs made use of deglitched frame files \citep{Nitz:2021zwj}. For the other events, there was no significant difference between the \ac{pe} runs with the old and new calibration methods.}
    \label{fig:calibration_differences}
\end{figure*}

For the event GW200208\_222617, there is strong bimodality in the mass ratio, $q$, for the GWTC \imrphenomxphm waveform model and the 4OGC analysis. In the baseline run, we reproduce the bimodality in mass ratio. This bimodality is still present in the APE parametrization, but it is no longer present in the RPE parametrization. There are hints of bimodality in the mass ratio for GW191109\_010717, with the GWTC analysis using the \imrphenomxphm waveform model showing a strong bimodality. We do not observe this bimodality in the waveform-uncertainty parametrization. For GW190521\_030229, there is a small bimodal peak at a low $q\sim0.2$ in the GWTC analysis using the \imrphenomxphm waveform model. We recover this small bimodal peak in baseline \ac{pe} run. The APE parametrization dilutes this small bimodal peak, but it remains. For RPE parametrization, this small peak disappears.

\section{Detector Calibration Uncertainties Correction}
\label{appendix:calibration_correction}

The strain data $d(t)$ obtained from the \ac{gw} detectors result from the detector's response to fluctuations in laser power at the photo detector, which are caused by relative differences in arm lengths. However, this strain data is not perfectly calibrated, and each detector is provided with an associated uncertainty error budget. These calibration uncertainties are accounted for in the \ac{pe} pipeline. As noted in \citep{Baka:2025bbb}, a problem was identified regarding the way calibration uncertainties were applied in earlier \ac{pe} analyses. In this analysis, the authors use a reweighting scheme applied to older posterior samples to correct for inaccurate calibration conventions. The authors reported that the error in the calibration convention does not significantly impact any conclusion of the previous analysis. 

In our \ac{pe} runs, when we use the corrected calibration scheme and confirm that we find a similar trend, i.e., there is no significant change in the posterior distribution except for GW191109. For this event, we point to figure \ref{fig:calibration_differences}, which shows the differences between posterior samples under the old and new calibration schemes across various scenarios.

\bibliographystyle{apsrev4-1}
\bibliography{references}
\end{document}